%% file: main.tex
\pdfoutput=1 
\documentclass[aps, prl, reprint, twocolumn, superscriptaddress, footinbib]{revtex4-1}
\usepackage[utf8]{inputenc}

\usepackage[colorlinks=true,urlcolor=blue,allcolors=blue,breaklinks=true]{hyperref}

\usepackage{tikz,tikz-3dplot}
\usetikzlibrary{patterns,calc,arrows,fadings}

\usepackage{color}
\definecolor{fred}{RGB}{228,26,28}
\definecolor{fblue}{RGB}{55,126,184}
\definecolor{fgreen}{RGB}{77,175,74}
\definecolor{fbrown}{RGB}{55,126,184}
\definecolor{fpurple}{RGB}{153,90,233}

\usepackage[normalem]{ulem}

\usepackage{pgfplots}
\usepackage{todonotes}
\usepackage{nicefrac}
\pgfplotsset{compat=newest}

\makeatletter 
\pgfdeclarepatternformonly[\LineSpace,\tikz@pattern@color]{my north east lines}{\pgfqpoint{-1pt}{-1pt}}{\pgfqpoint{\LineSpace}{\LineSpace}}{\pgfqpoint{\LineSpace}{\LineSpace}}%
{
	\pgfsetcolor{\tikz@pattern@color} 
	\pgfsetlinewidth{0.4pt}
	\pgfpathmoveto{\pgfqpoint{0pt}{0pt}}
	\pgfpathlineto{\pgfqpoint{\LineSpace + 0.1pt}{\LineSpace + 0.1pt}}
	\pgfusepath{stroke}
}
\makeatother 
\newdimen\LineSpace
\tikzset{
	line space/.code={\LineSpace=#1},
	line space=3pt
}

\usepackage{amsmath}
\usepackage{amssymb}
\usepackage{bbold}
\usepackage{bigdelim} 
\usepackage{array}

\newcommand{\ccd}{\operatorname{c}^\dagger}
\newcommand{\cc}{\operatorname{c}^{\vphantom{\dagger}}}
\newcommand{\Rd}{\operatorname{R}^\dagger}
\newcommand{\Rr}{\operatorname{R}^{\vphantom{\dagger}}}

\newcommand{\Ll}{\operatorname{L}^{\vphantom{\dagger}}}

\begin{document}

	\author{Florian St{\"a}bler}
	\affiliation{Ludwig Maximilian University, Arnold Sommerfeld Center and Center for Nano-Science, Munich, DE-80333, Germany}
        \author{Alexei M. Tsvelik}
        \affiliation{Brookhaven National Laboratory, Upton, NY 11973-5000, USA}
	\author{Oleg M. Yevtushenko}
	\affiliation{Ludwig Maximilian University, Arnold Sommerfeld Center and Center for Nano-Science, Munich, DE-80333, Germany}

        \title{Protected helical transport in magnetically doped quantum wires: \\
               Beyond the one-dimensional paradigm}
	\date{\today}

\begin{abstract}
One-dimensional (1D) quantum wires, which are functionalized
by magnetic ad-atoms, can host ballistic helical transport.
Helicity protects transport from an undesirable influence of
material imperfections, and it makes the magnetically doped wire
a very promising element for nanoelectronics and spintronics.
However, fabricating purely 1D conductors is experimentally
very challenging and not always feasible. In this paper, we
show that the protected helical transport can exist even in
quasi-1D wires. We model the quasi-1D magnetically doped wire
as two coupled dense 1D Kondo chains. Each chain consists
of itinerant electrons interacting with localized quantum
magnetic moments -- Kondo impurities. We have analyzed the
regimes of  weak-, intermediate-, and strong
inter-chain coupling, and we found conditions necessary for the
origin of the aforementioned protected transport.
Our results may pave the way for experimental realizations of
 helical states in magnetically doped wires.
\end{abstract}
	
\maketitle

One major stepping stone in the progress of nanoelectronics and spintronics is the reduction of destructive effects caused by material imperfections,
e.g. backscattering and localization. One-dimensional (1D) conductors are especially
sensitive to such undesirable effects that suppress ballistic transport \cite{Giamarchi}.
One possibility for protected transport is provided by the helicity of conduction electrons.
Helicity, \(h=\text{sgn}(p)\cdot\text{sgn}(\sigma)\), reflects the lock-in
relation between the electron's momentum, \(p\), and spin, \(\sigma\). Transport in a quantum wire is helical and, hence, protected when all conduction
electrons have the same helicity.

Physical mechanisms, which yield helical states,
generally fall into two categories. The first category includes topological
insulators (TI) \cite{TI3,TI1,topin}.
1D conducting helical modes can exist on edges of 2D TI
\cite{BenZhang,BenZhang2,AssymQ-Wells}.
%
%
Modern experiments show  helicity of hinge states in high-order TI \cite{BismuthTI-Meas,murani_ballistic_2017BI,schindler_higher-order_2018}.
Protection of the helical edge transport is expected to be ideal though
it is not robust in reality
\cite{EdgeTransport-Exp0,EdgeTransport-Exp1,EdgeTransport-Exp2,Koenig,Koenig2,vayrynen_2016,yud,oleg4,olegyud}.

The second category includes systems in which the helical states are governed by
interactions, e.g., the hyperfine interaction between nuclear magnetic
moments and itinerant electrons \cite{Loss,klin,loss2,CNTklinloss,CNTklinloss2}, and the
spin-orbit interaction (SOI) in a combination with either magnetic fields \cite{quay,heedt}
or Coulomb interactions \cite{Carr}. Several experiments
confirmed the existence of helical states in interacting systems \cite{zum,quay,heedt,Kammhuber}.

Another promising platform for protected helical transport is provided by magnetically
doped 1D quantum wires \cite{oleg1,schimmel1,oleg2,oleg3}. It is somewhat similar to the successful
realization of topological superconductivity \cite{Feldman2017,Desjardins2019,yaz}.
Despite the solid theoretical background,
experiments demonstrating helical transport in magnetically doped 1D wires
are still missing. The main obstacle hampering these experiments in traditional materials
(GaAs or SiGe) is the non-triviality of methods used to produce  1D conductors \cite{CEO,SiGe}.

{\it The goal of this Letter} is to show that the strict one-dimensionality is not
necessary and the protected helical states can emerge also in quasi-1D samples.

\begin{figure}[t]
	\input{figures/KC}
\vspace{-0.3 cm}
	\caption{The quasi-1D doped quantum wire is modeled by two coupled Kondo chains, each consisting
          of itinerant electrons (orange tubes) and an array of localized quantum magnetic impurities
          (blue spins). Electrons can tunnel at every site of the electronic lattice into the neighboring wire,
          tunneling is indicated by dashed lines.}
	\label{fig:KC}
  \vspace{-0.75 cm}
\end{figure}
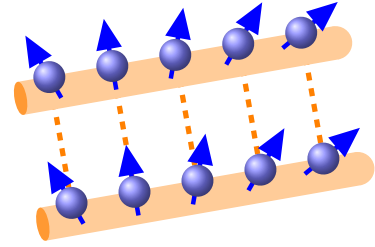

{\it Model}:
Magnetically doped quantum wires can be described by the well-known theoretical model
of a Kondo chain (KC) - a 1D array of localized quantum magnetic impurities
interacting with 1D itinerant electrons \cite{KC1,KC2,KC3,KC4,KC6,KC5,KC7,MC,zach}.
The physics of the KC is governed by two competing, mutually exclusive effects: the Kondo effect
and the indirect, Ruderman–Kittel–Kasuya–Yosida (RKKY), exchange
interaction between the impurities.
The dominant effect can be found from a comparison of relevant energy scales: the Kondo temperature, $ T_K $, and the RKKY energy, $ E_{\rm RKKY} $ \cite{doni}. If $ T_K > E_{\rm RKKY} $, the Kondo screening dominates; magnetic impurities are screened individually. This leads to a Kondo insulator at half-filling and a heavy fermion phase away from half-filling \cite{KC1,coleman}. In the opposite case $ T_K < E_{\rm RKKY} $, the RKKY interaction dominates and governs inter-impurity correlations. One can translate the above inequality to
distances and show that RKKY overwhelms the Kondo effect in dense KCs, where the (mean) inter-impurity distance \(\xi_s\) is smaller than a crossover scale \(\xi_c\): $ \xi_s \leq \xi_c \propto \xi_0 \left(\rho_0 J^2/T_K\right)^{1/2} $ \cite{oleg1,oleg2,oleg3,schimmel1}. Here \(J, \rho_0 \), and $ \xi_0 $ are the Kondo coupling, the density of states, and the lattice spacing, respectively. The RKKY-dominated regime is typical in 1D
systems \cite{oleg5,oleg2}.
 We have recently shown that helical spin ordering and
protected transport can exist in the dense and incommensurate KC with a small Kondo coupling,
which can be anisotropic (easy-plane anisotropy) \cite{oleg1,schimmel1} or isotropic \cite{oleg2,oleg3}.

To demonstrate, that helical transport can exist in quasi-1D wires,
we consider two tunneling-coupled dense and incommensurate KCs (Fig.\ref{fig:KC}). This simplest (minimal) quasi-1D model can provide a proof of principle since it
possesses a nontrivial degree of freedom: The magnetic impurities in different KCs are correlated only via tunneling, and it is a-priori not clear, whether they form a global helical ordering, which has the same handedness in each KC
\footnote{
We remind readers that, in the 1D case, spins form a helix in a transverse plane.
This is reflected by the spin susceptibility  $ \langle S^+_l S^-_m \rangle
\sim e^{\pm 2 i k_F \xi_s |l - m|}, S^\pm = S^x \pm i S^y $, which has either \(+2k_F\)
or  \(-2k_F\) component depending on the helix handedness \cite{schimmel1}}.
Such correlations can
protect ballistic transport in the quasi-1D system. The Hamiltonian of our quasi-1D model reads 
\begin{eqnarray}
\label{Ham}
   & \!\!\! & \hat{H} = \hat{H}_{KC}^{(1)} + \hat{H}_{KC}^{(2)} + \hat{H}_{\rm tun} \, ; \
   \hat{H}_{\rm tun}=- t_\perp c_{j}^{\dagger (1)} c_{j}^{(2)} + h.c.; \cr
   & \!\!\! & \hat{H}_{KC}^{(n)} = -t \left( \ccd_{j}\cc_{j+1} \right)^{(n)}+
                      J_a \left( \ccd_{j} S^a_{j} \operatorname{\sigma}_a \cc_{j} \right)^{(n)} + h.c. ;
\end{eqnarray}

where $ \hat{H}_{KC}^{(1,2)} $ are the Hamiltonians of the uncoupled KCs;
$ \hat{H}_{\rm tun}$ describes the electron tunneling;
\(c_j=\left(c_{j\uparrow},c_{j_\downarrow}\right)^T\) is a spinor;
\(\operatorname{c}_{j\sigma}\) ($ c_{j\sigma}^\dagger $) annihilates (creates) an electron
with spin \( \sigma=\uparrow,\downarrow\) at a lattice site \(j\) of a given chain marked
by \(n=1,2\); \(t \) $ (t_{\perp}) $ is the intra (inter) chain hopping
strength; \(J_a\) is the strength of the Kondo interaction in the \(a=x,y,z\) direction;
 $ \mathbf{S} $ is an impurity spin \(s\) operator;
and \(\sigma_a\) are the Pauli matrices \footnote{We implicitly assume summation over all repeated indices.}.
For simplicity, we do not distinguish lattice constants
\(\xi_0\) and \(\xi_s\), we assume that the individual KCs have the same parameters, and we focus on  zero
temperature, $ T \rightarrow 0 $. We explore the case of the easy-plane magnetic anisotropy
$ J_x=J_y=J \gg J_z \to 0 $ with a small coupling constant,
 $ s J  \ll t $, and incommensurate band fillings. This setup is relevant for the search of  protected transport \cite{oleg1,schimmel1}
and much simpler for the theoretical study than the isotropic case \cite{oleg2,oleg3}.
Note that Kondo-like renormalizations are suppressed and can be neglected in the dense KCs
whose physics is dominated by the RKKY interaction \cite{schimmel1}.

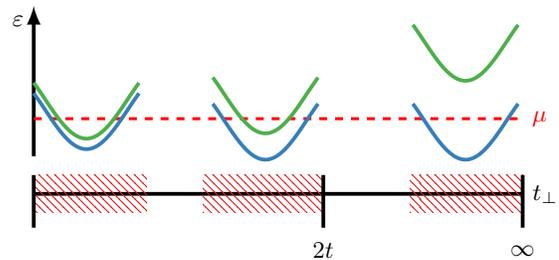
\begin{figure}[t]
\centering
\input{figures/threecases.tex}
\vspace{-0.3 cm}
\caption{Three regimes of
  inter chain tunneling (marked by shaded areas).
  Left panel: tunneling is weak, \(t_\perp \ll 2t\), and one comes across four Fermi points
  that almost coincide in pairs. Central panel: tunneling is larger, \(t_\perp \lesssim 2t \), and
   all four Fermi points are well separated.
  Right panel: tunneling is strong \(2t \ll t_\perp\), and energy bands are separated by the
  gap and there are at most two Fermi points
  }
  \vspace{-0.5 cm}
\label{fig:threecase}
\end{figure}

{\it Three regimes of the tunneling-coupled KC}:
The non-interacting part of the Hamiltonian (\ref{Ham}), $ \hat{H}_0 \equiv \hat{H}|_{J=0} $,
has the spectrum \( \varepsilon_\pm (k)=
-2t \cos\left(k\xi_0\right) \mp t_\perp-\mu\) \footnote{The new band operators \(\operatorname{c}_\pm=\frac{1}{\sqrt{2}}\left(\operatorname{c}_1 \pm \operatorname{c}_2\right)\) are the (anti-) symmetric linear combinations of the old operators. The lower + band is thus accompanied by a downward shift in energy \(\varepsilon_+=\varepsilon_0-t_\perp\) and vice versa. }.
The value of \(t_\perp\) determines three different regimes: the strong-, intermediate-, and weak- interchain tunneling; see Fig.\ref{fig:threecase}.

If tunneling is strong, $ 2t\ll t_\perp $, there are two bands separated by a large gap
of order $ t_\perp $. Without loss of generality, we can place the chemical potential, $ \mu $, in the lower
band and take into account the electron-spin interaction perturbatively by using the smallness
 \( s J / t_\perp \ll t / t_\perp \ll 1 \).
We will show that such a perturbation yields only
small and inessential corrections to the physics of the helical 1D wire
described in Refs.\cite{oleg1,schimmel1}.

The other two cases of the intermediate, \(t_\perp \lesssim 2t \), or small,
\(t_\perp \ll 2t\), tunneling,
can possess four Fermi points.
{\color{purple} }
In the former case,
the Fermi points are well separated and one has to take into account all electron-spin
interactions non-perturbatively. If tunneling is weak, the four Fermi points almost coincide
in pairs,
 and small \(t_\perp\) can be treated as a perturbation for two decoupled KCs.

We rewrite the Kondo interaction in the eigenbasis of $ \hat{H}_0 $:
\begin{multline}\label{eq:lagblin}
  \hat{H}_{\rm int}=
  (J / 2) \left[
    \ccd_\nu S_{+}^b \operatorname{\sigma}_b \cc_\nu +
    \ccd_\nu S_{-}^b \operatorname{\sigma}_b \cc_{-\nu} + h.c.
          \right],
\end{multline}
where \(b=x,y\), \(\mathbf{S}_\pm= \mathbf{S}^{(1)}\pm \mathbf{S}^{(2)}\) and \(\nu = +(-)\)
labels the lower (upper) band. The Kondo interaction enables intra- and interband scatterings.
We  will use the functional integral formulation of the theory on the imaginary time contour
and analyze the three cases shown in Fig.~\ref{fig:threecase}. The localized spins in this
approach are conveniently parameterized  by a  normalized vector field \cite{tsvelikbook1}.

{\it{Strong tunneling,  \(J\ll2t\ll t_\perp\)}}:
If $ \mu $ belongs to the lower band and $ T = 0 $, transitions
between the bands are virtual and result only in a small renormalization of parameters of the
conduction band \footnote{The case where the chemical potential belongs to the upper band can
be treated analogously. }. To show this, we integrate out the fermions from the upper band perturbatively.
This yields a mass term for the propagator of the conduction electrons from the lower band:
$ \Sigma_{-} = J^2 \left({ S}^b_-\right)^2 \langle \psi_- \bar{\psi}_- \rangle \simeq - (J^2 / 2t_\perp)
\left({ S}^b_-\right)^2 + \mathcal{O}(J^2/t_\perp^2) $ \cite{SM}; $ \psi_{\pm} $
are fermionic fields. $ \Sigma_{-} $ governs
a shift of $ \mu $ and enables a weak spin conserving backscattering. Both effects are parametrically
small compared to those governed by the intraband Kondo interaction. Therefore, the interband
transitions can be neglected and the Lagrangian density of the electrons in the lower band reduces to
\begin{equation}
\label{Lagr-ST}
  \mathcal{L}_+^{(\rm ST)} \simeq \bar{\psi}_+ \left[-i\omega+\epsilon_+(k) -\mu +
    ( J \rho_s / 2 ) \, S_{+}^b \operatorname{\sigma}_{b}
                                  \right] \psi_+;
\end{equation}
where $ \omega $ is the fermionic Matsubara frequency, and $ \rho_s $ is the spins density. Below,
we will change to the continuous limit with $ \rho_s = {\rm const} $ and absorb $ \rho_s $ in the
coupling constant: $ J' \equiv J \rho_s / 2 $. Eq.(\ref{Lagr-ST}) describes
a single KC where the itinerant electrons interact with
the composite spins $ \mathbf{S}_{+} $. This theory can be studied by using the approach
developed in Refs.\cite{oleg1,schimmel1} for  1D KC. It can be straightforwardly
proven that model (\ref{Lagr-ST}) supports protected helical transport.

{\it{Intermediate tunneling, \(t_\perp \lesssim 2t\)}}: Let us analyze the case in which four Fermi points
(two in the lower band and two in the upper band with Fermi momenta  \(\pm k_{F}^{(\pm)} \), respectively)
coexist and are well separated, $ \delta k_F \equiv k_{F}^{(+)} - k_{F}^{(-)} \sim \tilde{k}_F \equiv
( k_{F}^{(+)} + k_{F}^{(-)} ) /2 $.
 We have to single out slow modes.
We linearize the dispersion relation of the non-interacting system around the Fermi points and
introduce smooth left ($\rm L $) and right ($\rm R $) moving modes in a standard way.
These fermionic modes are described by the Lagrangian $ \mathcal{L}_0 = \bar{\rm R}_\nu \partial_{+}^{(\nu)}
\rm{R}_\nu + \bar{\rm L}_\nu \partial_{-}^{(\nu)} \rm{L}_\nu $ with $ \partial_{\pm}^{(\nu)} = \partial_\tau
\mp i v_F^{(\nu)} \partial_x $ being the chiral derivative. The Fermi velocity depends on the band
index: \(v_{F}^{(\nu)}= 2t\xi_0 \sin\left(k_{F}^{(\nu)}\xi_0 \right)\).

The physics of the dense KCs is governed by backscattering of the fermions \cite{oleg1,schimmel1,oleg2,oleg3}
described by
\begin{equation}
\label{eq:bsnu1nu2}
   \mathcal{L}_{\rm bs}^{\nu\nu'} = J' \Rd_\nu S^b_{\pm} \sigma_b \Ll_{\nu'} e^{2i k_{F}^{\nu\nu'}x} + h.c.
\end{equation}
$ k_{F}^{\nu\nu'} = \tilde{k}_F $  for the interband backscattering \((S_-)\),
$ \nu = - \nu' $, and $ k_{F}^{\nu\nu'} =  \tilde{k}_F + \nu \delta k_F / 2 $   for the intraband one \((S_+)\),
$ \nu =  \nu' $. Backscattering opens a gap in the spectrum of fermions and, thus, reduces the ground
state  (GS) energy of the entire system \cite{oleg1,oleg2,oleg3,schimmel1}.

We are interested in the low-energy physics whose Lagrangian does not contain $ 2 k_F $-oscillations.
Our strategy is to absorb them into spin configurations and find the configuration, that minimizes
the GS energy by maximizing backscattering. We decompose the spin variables into
slow and fast components \cite{SM}:
\begin{equation}
\label{SpinParam}
  {\bf S}^{(n)}/s  = \mathbf{m}_n\! + \! \left[\mathbf{e}_{1}^{(n)} \!\cos\left(Q x
    \right)+\mathbf{e}_{2}^{(n)} \!\sin\left(Q x \right) \right] \! \sqrt{1-\mathbf{m}_n^2} .
\nonumber
\end{equation}
Here $ \mathbf{m}_n= \sin\left(\alpha^{(n)}\right) [ \mathbf{e}_{1}^{(n)} \times \mathbf{e}_{2}^{(n)} ] $,
$ 2 k_F^{(-)} \le Q \le 2 k_F^{(+)} $ and $ \mathbf{e}_{1,2}^{(n)} $ are two orthonormal vectors that lie
almost in the plane defined by the magnetic anisotropy (``easy plane''). These two vectors
are
parameterized by the in-plane polar angle, $ \psi^{(n)} $, and by another angle describing small
out-of-plane fluctuations, $ \theta^{(n)} $. Oscillating terms allow one
to absorb $ 2 k_F $-oscillations from the backscattering and, thus, are needed to minimize the
GS energy. The angle $ \alpha^{(n)} $ weighs the zero mode \(\mathbf{m}_n\) and has the semiclassical value $ \alpha_{\rm cl}^{(n)} \to 0 $. Deviations of $ \alpha^{(n)} $ from this value are small.
$ \theta^{(n)} $ and $ \alpha^{(n)} $ are massive variables and they can be integrated out in the Gaussian
approximation \cite{oleg1,schimmel1}.

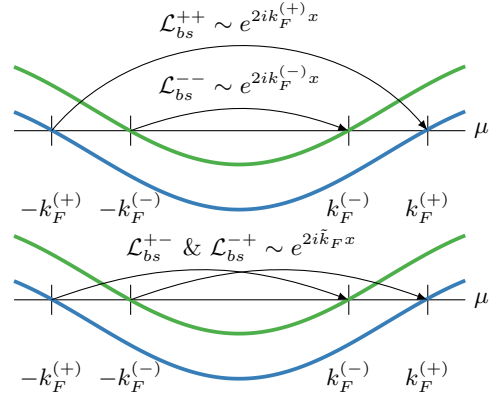
\begin{figure}[t]
	\input{figures/OSC}
\vspace{-0.3 cm}

	\caption{Upper/Lower panels: intraband/interband scattering processes and corresponding oscillating
             factors in Eq.(\ref{eq:bsnu1nu2}).
             }
\vspace{-0.5cm}
	\label{fig:osc}
\end{figure}

Eq.(\ref{eq:bsnu1nu2}) contains oscillations with three different wave vectors, $ 2 k_F^{(+)} = 2 k_F^{(++)},
2 k_F^{(-)} = 2 k_F^{(--)}, \mbox{ and } 2 \tilde{k}_F = 2 | k_F^{(+-)} | $, which are of the same order in the
intermediate tunneling regime and correspond to various intra- and inter-band scatterings, see Fig.\ref{fig:osc}.
By tuning $ Q $, one can
absorb into the spin configuration only one of these
vectors; the other two result in fast oscillations
that do not contribute to the low-energy theory. The remaining smooth part of the backscattering opens
the helical gap [see Eq.(\ref{eq:helgap1}) below] in the fermionic spectrum.
If $ Q = 2 k_F^{(\pm)} $, the gap is opened only in one  (either lower or upper) band.
The choice \(Q=2\tilde{k}_F\) results in doubling the number of  gapped fermionic modes.
Moreover, it provides the maximal value of all gaps \cite{SM}. We
thus conclude that the GS energy reaches its minimum at \(Q=2\tilde{k}_F\). After inserting this
choice into Eq.(\ref{eq:bsnu1nu2}) and neglecting oscillating terms, we arrive at:
\begin{align}\label{eq:afternunu'}
  \mathcal{L}_{bs}^{\nu\nu'}& \simeq
               \Rd_{\nu}\left[ \hat{\Delta}^{(1)}- \hat{\Delta}^{(2)}\right] \Ll_{\nu'} + h.c., \quad \nu \neq \nu',
\end{align}
where \( \hat{\Delta}^{(n)}\) are scattering amplitudes of the respective 1D KCs \cite{schimmel1,oleg1,oleg2,oleg3}:
$ \hat{\Delta}^{(n)} / \tilde{J} = e^{i\psi^{(n)}}\!\sin^2\left( \theta^{(n)} / 2 \right)  \sigma_- -
e^{-i\psi^{(n)}} \!\cos^2 \left( \theta^{(n)} / 2 \right) \sigma_+; \,
\tilde{J}=s \cos(\alpha^{(n)}) J' $. Next, we use the classical value \(\alpha^{(n)}_{\rm cl}=0\)
and look for the classical value of $ \theta^{(n)} $. We anticipate that $ \theta^{(n)}_{\rm cl} = 0 \mbox{ or } \pi $, \cite{oleg1,schimmel1}.

The
gap values (at fixed angles $ \psi^{(n)} $) are different in the cases $ \theta^{(1)}
= \theta^{(2)} $ and $ \theta^{(1)} \ne \theta^{(2)} $. For example, if $ \theta^{(1)} = 0 $ then
\begin{eqnarray}
\label{eq:helgap1}
  \theta^{(2)} = 0: & \quad & \hat{m}_- = 2i \tilde{J} e^{-i \tilde{\psi}} \! \sin\left(\delta\psi\right)\sigma_+ , \\
\label{eq:helgap2}
  \theta^{(2)} = \pi: & \quad & \hat{m}_- = - \tilde{J}
                                             \left(e^{i\psi^{(2)}} \! \sigma_- + e^{-i\psi^{(1)}} \! \sigma_+ \right).
\end{eqnarray}
Here \(\hat{m}_- \equiv \hat{\Delta}^{(1)}-\hat{\Delta}^{(2)}\); \(\tilde{\psi} \equiv \left(\psi^{(1)}+\psi^{(2)}\right)/2\);
and \(\delta\psi \equiv (\psi^{(1)}-\psi^{(2)}) / 2 \). The modulus of the eigenvalue of $ \hat{m} $ reaches maximum in
Eq.(\ref{eq:helgap1}) at $ \delta\psi = \pm \pi / 2 $ and becomes twice as large as that in Eq.(\ref{eq:helgap2}). Therefore,
we come across a mode locking of the in-plane spin polar angles which makes the spin configuration $ \theta^{(1)} = \theta^{(2)} $
energetically favorable \cite{SM}.
The phase factor \(\tilde{\psi}\) in Eq.(\ref{eq:helgap1}) can be gauged out.
 This leads to
the expression for the gain (with respect to the non-interacting case, $ J = 0 $) of the GS energy \cite{SM}:
\begin{equation}
\delta E^{\rm (IT)} = - \left[4 \xi_0 \tilde{J}^2 / \pi \left(v_{F_+}+v_{F_-}\right)\right]
                           \ln\left( 2t / |\tilde{J}| \right);
\end{equation}
 The analysis of GS shows, that the helical symmetry is spontaneously broken and a gap opens for fermions with a given helicity in both bands. As a result, we find gapless helical fermions with \(h=-1\) for \(\theta^{(1,2)}=0\) (or \(h=+1\) for $\theta^{(1,2)}=\pi$).

To finalize the derivation of the effective low-energy theory, we reinstate the Wess-Zumino term for the spin variables
\cite{tsvelikbook1} and integrate out all massive fields approximately \cite{SM}.
This yields the Lagrangian

$
	\mathcal{L}^{\rm (IT)}= \mathcal{L}_{\rm LL} [\tilde{\psi}] + \sum_{\nu = \pm} \mathcal{L}_0[\rm R_{\nu\downarrow} , L_{\nu\uparrow}] .
$

Here \(\mathcal{L}_{\rm LL} = [ (\partial_\tau \tilde{\psi})^2 + (v_\psi \partial_x \tilde{\psi})^2 ] / 2 \pi K_\psi \) is the Luttinger
liquid Lagrangian, which describes the slow, $ v_\psi \ll v_F $, collective bosonic helical mode with the effective strong interactions,
$ K_\psi \ll 1 $.
Gapless fermionic modes have the same helicity in each band, $ \nu = \pm $.
This parametrically suppresses Anderson localization which can be induced by an additional  spinless disorder, with the disorder strength being $ < J $ \cite{SM}. Thus, transport in these systems is
protected by the helicity and remains ballistic in parametrically long samples.

{\it{Weak tunneling, \(t_\perp \leq J \ll 2t\):}}  If \(t_\perp\) is small, the separation between the Fermi
points shrinks and they almost coincide in pairs when \( \delta k_F \approx 2t_\perp / \tilde{v}_F \ll \tilde{k}_F \).
We start again from Eq.(\ref{eq:bsnu1nu2}), however, unlike the intermediate tunneling,
$ \delta k_F $-oscillations are slow and cannot be neglected in the low energy sector. This makes the number of the gapped
fermionic modes independent of the choice of \(Q\). We retain \(Q=2\tilde{k}_F\) for convenience and repeat the
steps resulting in Eq.(\ref{eq:afternunu'}). Slow $ \delta k_F $-oscillations yield now additional intraband scattering terms:
\begin{align}\label{eq:afternunu}
\mathcal{L}_{bs}^{\nu\nu}& \simeq
\Rd_{\nu}\left[ \hat{\Delta}^{(1)}+ \hat{\Delta}^{(2)}\right] \Ll_{\nu} e^{i\nu\delta k_F x} + h.c.
\end{align}
The slowly oscillating backscattering opens a gap at the energy which is shifted by $ \delta k_F v_F $ from $ \mu $,
leading to a small number of occupied (or empty) states above (or below)
the gap \cite{oleg2,oleg3}. These states are energetically split off by the gap and thus have no noticeable influence on the dc transport.

Next, we use the value \(\alpha_{\rm cl}^{(n)}=0\) and look for the optimal spin configuration with
\(\theta_{\rm cl}^{(n)}=0\) or \(\pi\). The intraband scattering introduces a new gap structure. Additionally to
Eqs.(\ref{eq:helgap1},\ref{eq:helgap2}), we find for $ \theta^{(1)} = 0 $:
\begin{align}\label{eq:helgap1+}
  \theta^{(2)} = 0 &: \quad \hat{m}_+ = -2 \tilde{J} e^{-i \tilde{\psi}} \cos\left(\delta\psi\right)\sigma_+ ,        \\
  \label{eq:helgap2+}
  \theta^{(2)} = \pi &: \quad \hat{m}_+ = \tilde{J} \left(e^{i\psi^{(2)}} \sigma_--e^{-i\psi^{(1)}} \sigma_+\right);
\end{align}
where \(\hat{m}_+ =\hat{\Delta}^{(1)}+\hat{\Delta}^{(2)}\). We integrate out the gapped fermions, expand the
result perturbatively in \( t_\perp / t \ll 1 \), and find the
expression for the (relative) GS energy of the weakly coupled KCs \cite{SM}:
\begin{eqnarray}
\label{Ewt}
  & & \!\!\!
  \delta E^{\rm (WT)} \approx - (2\xi_0 \tilde{J}^2 / \pi \tilde{v}_F) \ln\left( 2t / |\tilde{J}|\right)
  \Bigl\{ 1 + \\
%
  & &
  \left[t_\perp / t \, \sin\left(\tilde{k}_F \xi_0\right) \right]^2 \left( 1 + 2\cos^2\left(\delta\psi\right)\cot^2\left(\tilde{k}_F \xi_0\right)\right)\Bigr\},
  \nonumber
\end{eqnarray}
with  \(\tilde{v}_F=v_F(\tilde{k}_F)\) and  $ \xi_0 t_\perp / \tilde{v}_F \ll 1 $ .
The energy gain due to the mode locking, $ \delta \psi
\simeq 0 $ or $ \pi $, manifests itself in Eq.(\ref{Ewt}) starting from the term $ O(t_\perp / t)^2 $
and guarantees that the helical phase provides the minimum of the GS energy. The low energy theory
 is described by  $ {\cal L}^{\rm (IT)}$ with
 \(v_{F_+}=v_{F_-}=\tilde{v}_F\). We conclude that  transport is  helical and protected in weakly coupled KCs.

%

{\it Vanishing tunneling}:
If \(t_\perp \ll J \ll 2t\), the perturbative corrections to the GS energy in
Eq.(\ref{Ewt}) become beyond the accuracy of calculations. If one naively neglects them,
our model is reduced to two uncoupled 1D KCs whose GS is degenerate, either
\(\theta^{(1)}=\theta^{(2)}=0\) or \(\theta^{(1)} = 0, \, \theta^{(2)} = \pi\). The latter
configuration corresponds to the phase where gapless fermions have opposite helicity in
different wires. Clearly, two channels with opposite helicity form a usual (non-helical)
spinful conducting channel where transport is not protected.    However, this artificial
degeneracy does not mean violation of the helcial protection which can be reinstated
via the cumbersome analysis with higher accuracy.
We prefer to avoid unnecessary technical complications. To this end, we note that even a weak intrinsic Dresselhaus
SOI \cite{dresselhaus}, which typically exists in GaAs quantum wires, removes this
ambiguity and generates corrections to $ \delta E^{\rm (WT)} $ which again
drive the system to the helical phase with protected transport \cite{SM}.

{\it Conclusions}: We have shown that strict one-dimensionality is not
a necessary prerequisite for the formation of a helical phase with protected
transport in nanowires functionalized by magnetic adatoms. To demonstrate this
statement, we have studied the simplest theoretical model of two
dense magnetically anisotropic 1D Kondo chains coupled by the
interchain tunneling of itinerant electrons. The anisotropy simplifies calculations, however, preliminary analysis shows that our conclusions remain valid also in the isotropic case.
The ground state of  our model is manifestly helical when the interchain tunneling is
larger- or of the order of the exchange coupling between the itinerant electrons
and localized spins. The latter, in turn, must be much smaller than the width of
the conduction band,  but much larger than the temperature, $ T \ll J \ll t $.  These conditions are natural for experimental setups  where $ J $ can be tuned by using various magnetic ad-atoms and changing their density and proximity to the quantum wire. Small fluctuations of the Kondo couplings  \(J_{1,2} = \tilde{J} \pm \delta J, \ \delta J/\tilde{J}\ll 1\) cannot change our conclusions \cite{SM}.

The global helicity is provided by the indirect (intra- and interwire) interaction
between the localized spins. Adding more chains to the model can make the spin interaction
weaker, when the system approaches the 2D limit, but cannot violate the helical protection
in the quasi-1D samples. Our predictions are also stable with respect to a weak
or moderate Coulomb interaction of the electrons; cf. Ref.\cite{schimmel1}: the electrostatic
repulsion enhances the RKKY interaction and, thus, can only make interspin correlations
and the helical protection of transport stronger.
Thus, our results substantially expand predictions made for purely 1D wires \cite{schimmel1,oleg1},
and they could facilitate  experimental  studies of
protected transport in various magnetically doped nanostructures.

\begin{acknowledgments}
	 {\it Acknowledgements}: A.M.T. was supported by the Office of Basic Energy Sciences, Material Sciences and Engineering Division, U.S. Department of Energy (DOE) under Contract No. DE-SC0012704. O.M.Ye.  acknowledges support from the DFG through the grant YE  157/2-2. A.M.T. also acknowledges the hospitality of the Department of Physics of LMU.
\end{acknowledgments}

\bibliographystyle{apsrev4-1}


%


\input{SupplMat.tex}

\end{document}

%% file: figures/KC.tex
\begin{tikzpicture}

\begin{scope}[rotate=10,scale=.85]

\draw[dashed,orange,line width = 0.7mm] (0,0) -- (0,2);
\draw[dashed,orange,line width = 0.7mm] (1,0) -- (1,2);
\draw[dashed,orange,line width = 0.7mm] (2,0) -- (2,2);
\draw[dashed,orange,line width = 0.7mm] (3,0) -- (3,2);
\draw[dashed,orange,line width = 0.7mm] (4,0) -- (4,2);

\begin{scope}[rotate=0,shift={(0,2)}]
\begin{scope}[rotate=0,scale=0.5,shift={(-1,-2)}]
\filldraw[semithick,orange!40] (0,1) -- (10,1)--(10,2)--(0,2);
\filldraw[semithick,orange!40] (10,2) arc (90:-90:0.5 and 0.5);
\filldraw[semithick,orange!80] (0,1.5) ellipse (0.166 and 0.5);
\end{scope}

\begin{scope}[scale=.75,rotate around={20:(0,0)}]
\draw[line width= 0.6mm,blue,-triangle 45] (0,-0.5)--(0,1);
\end{scope}
\begin{scope}[scale=.75,rotate around={0:(1.33,0)}]
\draw[line width= 0.6mm,blue,-triangle 45] (1.33,-0.5)--(1.33,1);
\end{scope}
\begin{scope}[scale=.75,rotate around={-20:(2.66,0)}]
\draw[line width= 0.6mm,blue,-triangle 45] (2.66,-0.5)--(2.66,1);
\end{scope}
\begin{scope}[scale=.75,rotate around={-40:(4,0)}]
\draw[line width= 0.6mm,blue,-triangle 45] (4,-0.5)--(4,1);
\end{scope}
\begin{scope}[scale=.75,rotate around={-60:(5.33,0)}]
\draw[line width= 0.6mm,blue,-triangle 45] (5.33,-0.5)--(5.33,1);
\end{scope}

\shade[ball color=blue!50] (0,0) circle (.25);
\begin{scope}[shift={(1,0)}]
\shade[ball color=blue!50] (0,0) circle (.25);
\end{scope}
\begin{scope}[shift={(2,0)}]
\shade[ball color=blue!50] (0,0) circle (.25);
\end{scope}
\begin{scope}[shift={(3,0)}]
\shade[ball color=blue!50] (0,0) circle (.25);
\end{scope}
\begin{scope}[shift={(4,0)}]
\shade[ball color=blue!50] (0,0) circle (.25);
\end{scope}
\end{scope}

\begin{scope}[rotate=0]
\begin{scope}[rotate=0,scale=0.5,shift={(-1,-2)}]
\filldraw[semithick,orange!40] (0,1) -- (10,1)--(10,2)--(0,2);
\filldraw[semithick,orange!40] (10,2) arc (90:-90:0.5 and 0.5);
\filldraw[semithick,orange!80] (0,1.5) ellipse (0.166 and 0.5);
\end{scope}

\begin{scope}[scale=.75,rotate around={20:(0,0)}]
\draw[line width= 0.6mm,blue,-triangle 45] (0,-0.5)--(0,1);
\end{scope}
\begin{scope}[scale=.75,rotate around={0:(1.33,0)}]
\draw[line width= 0.6mm,blue,-triangle 45] (1.33,-0.5)--(1.33,1);
\end{scope}
\begin{scope}[scale=.75,rotate around={-20:(2.66,0)}]
\draw[line width= 0.6mm,blue,-triangle 45] (2.66,-0.5)--(2.66,1);
\end{scope}
\begin{scope}[scale=.75,rotate around={-40:(4,0)}]
\draw[line width= 0.6mm,blue,-triangle 45] (4,-0.5)--(4,1);
\end{scope}
\begin{scope}[scale=.75,rotate around={-60:(5.33,0)}]
\draw[line width= 0.6mm,blue,-triangle 45] (5.33,-0.5)--(5.33,1);
\end{scope}

\shade[ball color=blue!50] (0,0) circle (.25);
\begin{scope}[shift={(1,0)}]
\shade[ball color=blue!50] (0,0) circle (.25);
\end{scope}
\begin{scope}[shift={(2,0)}]
\shade[ball color=blue!50] (0,0) circle (.25);
\end{scope}
\begin{scope}[shift={(3,0)}]
\shade[ball color=blue!50] (0,0) circle (.25);
\end{scope}
\begin{scope}[shift={(4,0)}]
\shade[ball color=blue!50] (0,0) circle (.25);
\end{scope}
\end{scope}

\end{scope}

\end{tikzpicture}

%% file: figures/threecases.tex
\begin{tikzpicture}
 \draw[ line width=.5mm] (0,0) -- (6.5,0) node[right] {$t_\perp$};
 \draw[pattern=north west lines, pattern color=fred,
 draw=none] (0,-.25) rectangle (1.5,.25);
 \draw[pattern=north west lines, pattern color=fred,
 draw=none] (2.25,-.25) rectangle (3.85,.25);
 \draw[pattern=north west lines, pattern color=fred,
 draw=none] (5,-.25) rectangle (6.5,.25);
\draw[line width=.5mm] (0,.25)--(0,-.45);
\draw[line width=.5mm] (6.5,.25)--(6.5,-.45);
\draw[line width=.5mm] (3.85,.25)--(3.85,-.45);

\node (twot) at (3.85,-.75) {$2t$};
\node (infty) at (6.5,-.75) {$\infty$};

\draw[-latex,line width=0.5mm] (0,.5) -- (0,2.5) node[anchor= north east] {$\varepsilon$};
\draw[dashed,red,line width=0.4mm] (0,1) -- (6.5,1) node[right] {$\mu$};

\begin{scope}[scale=.7,shift={(1,1.7)}]
\draw[scale=1,domain=-1:1,smooth,variable=\x,fblue
,line width=0.5mm, samples=360] plot (\x,{-.75*cos(2*\x r)-.1}) node[right] {};
\draw[scale=1,domain=-1:1,smooth,variable=\x,fgreen
,line width=0.5mm, samples=360] plot (\x,{-.75*cos(2*\x r)+.1}) node[right] {};
\end{scope}

\begin{scope}[scale=.7,shift={(4.4,1.7)}]
\draw[scale=1,domain=-1:1,smooth,variable=\x,fblue
,line width=0.5mm, samples=360] plot (\x,{-.75*cos(2*\x r)-.3}) node[right] {};
\draw[scale=1,domain=-1:1,smooth,variable=\x,fgreen
,line width=0.5mm, samples=360] plot (\x,{-.75*cos(2*\x r)+.2}) node[right] {};
\end{scope}

\begin{scope}[scale=.7,shift={(8.2,1.7)}]
\draw[scale=1,domain=-1:1,smooth,variable=\x,fblue
,line width=0.5mm, samples=360] plot (\x,{-.75*cos(2*\x r)-.3}) node[right] {};
\draw[scale=1,domain=-1:1,smooth,variable=\x,fgreen
,line width=0.5mm, samples=360] plot (\x,{-.75*cos(2*\x r)+1.2}) node[right] {};
\end{scope}

\end{tikzpicture}

%% file: figures/OSC.tex
\centering

	\begin{tikzpicture}[scale=1]
	\draw (0.5,0) -- (6.5,0) node[right] {$\mu$};
	\draw (1,.2) -- (1,-.2) node[below,shift={(0,-.5)}] {$-k_{F}^{(+)}$};
	\draw (6,.2) -- (6,-.2) node[below,shift={(0,-.5)}] {$k_{F}^{(+)}$};
	\draw (2.05,.2) -- (2.05,-.2) node[below,shift={(0,-.5)}] {$-k_{F}^{(-)}$};
	\draw (4.95,.2) -- (4.95,-.2) node[below,shift={(0,-.5)}] {$k_{F}^{(-)}$};
	\draw[scale=1,domain=0.5:6.5,smooth,variable=\x,fgreen
	,line width=0.5mm, samples=360,opacity=1] plot (\x,{-.75*cos(4/5*\x r + 200)+.3}) node[right] {};
	\draw[scale=1,domain=0.5:6.5,smooth,variable=\x,fblue
	,line width=0.5mm, samples=360,opacity=1] plot (\x,{-.75*cos(4/5*\x r + 200)-.3}) node[right] {};
	\draw[-latex](1,0) to [out=50,in=130] node[ midway, above]{$\mathcal{L}_{bs}^{++} \sim e^{2ik_F^{(+)}x} $} (6,0);
	\draw[-latex](2.05,0) to [out=20,in=160] node[ midway, above]{$\mathcal{L}_{bs}^{--} \sim e^{2ik_F^{(-)}x} $} (4.95,0);
		\begin{scope}[shift={(0,-2.25)}]
		\draw (0.5,0) -- (6.5,0) node[right] {$\mu$};
	\draw (1,.2) -- (1,-.2) node[below,shift={(0,-.5)}] {$-k_{F}^{(+)}$};
	\draw (6,.2) -- (6,-.2) node[below,shift={(0,-.5)}] {$k_{F}^{(+)}$};
	\draw (2.05,.2) -- (2.05,-.2) node[below,shift={(0,-.5)}] {$-k_{F}^{(-)}$};
	\draw (4.95,.2) -- (4.95,-.2) node[below,shift={(0,-.5)}] {$k_{F}^{(-)}$};
	\draw[scale=1,domain=0.5:6.5,smooth,variable=\x,fgreen
	,line width=0.5mm, samples=360,opacity=1] plot (\x,{-.75*cos(4/5*\x r + 200)+.3}) node[right] {};
	\draw[scale=1,domain=0.5:6.5,smooth,variable=\x,fblue
	,line width=0.5mm, samples=360,opacity=1] plot (\x,{-.75*cos(4/5*\x r + 200)-.3}) node[right] {};
	\draw[-latex](1,0) to [out=20,in=160] node[ midway, above, shift={(0.55,0)}]{$\mathcal{L}_{bs}^{+-} \text{ \& } \mathcal{L}_{bs}^{-+} \sim e^{2i\tilde{k}_F x} $} (4.95,0);
	\draw[-latex](2.05,0) to [out=20,in=160] (6,0);
		\end{scope}
	\end{tikzpicture}

%% file: SupplMat.tex
\widetext

\newpage

\begin{center}
  {\large
    {\bf
   Supplemental Materials for the manuscript \\ \vspace{0.25cm}
    }
   "Protected helical transport in magnetically doped quantum wires: beyond the 1D paradigm" \\ \vspace{0.25cm}
  }
   by F. St{\"a}bler, A.M. Tsvelik, and O.M. Yevtushenko
\end{center}

\appendix


\section{Suppl.Mat. A:  Derivation of the effective Lagrangian in the strong tunneling limit}

Let us consider two strongly coupled 1D KCs, \(2t\ll t_\perp\). The non-interacting band structure of the system is shown in the right panel of Fig. \(\ref{fig:threecase}\). Our goal is to integrate out the upper band fermions to obtain an effective action for the lower band fermions, where we place the chemical potential. Next, we switch on a finite, but weak exchange interaction of the impurity spins and the conduction electrons \(J \ll 2t\).
We define the Green's function of the lower and upper band.

\begin{align}
-\mathbf{G}^{-1}_+&= -i\omega+\epsilon_0(k)-\mu' + J \boldsymbol{\mathcal{M}}_+,\\
-\mathbf{G}^{-1}_-&= -i\omega+\epsilon_0(k)- \mu' +2t_\perp+ J \boldsymbol{\mathcal{M}}_+,
\end{align} where we redefined redefined \(\mu=\mu'-t_\perp\) and \(\varepsilon_0(k)=\varepsilon_\pm(k)|_{t_\perp=0}\) and \(\boldsymbol{\mathcal{M}}_\pm= \sum_{b=x,y} \frac{1}{2}\mathbf{S}_{\pm}^b \operatorname{\sigma}_{b}\) is a matrix, which contains all the backscattering amplitudes generated by the spin \(\mathbf{S}_\pm\). We integrate out the upper band (\(-\)) fermions using the identity

\begin{equation}\label{eq:strongintout}
\left\langle \exp\left\{J\int \mathop{d\zeta} \ccd_+ \boldsymbol{\mathcal{M}}_- \cc_- + h.c.\right\} \right\rangle_- =\exp\left\{J^2\int \operatorname{d\zeta_{1,2}}  \ccd_+ \left.\boldsymbol{\mathcal{M}}_{-}\vphantom{\boldsymbol{\mathcal{M}}_{-}^T} \right|_{\zeta_1} \left.\mathbf{G}_- \vphantom{\boldsymbol{\mathcal{M}}_{-}^T}\right|_{\zeta_1-\zeta_2}  \left.\hspace{-.1cm}\vphantom{\boldsymbol{\mathcal{M}}_{-}^T}\boldsymbol{\mathcal{M}}_{-}\right|_{\zeta_2} \cc_+ \right\}
\end{equation} where \(\zeta_n=\{\tau_n,x_n\}\). In the next step, we use the smallness of \(\frac{J}{t_\perp}\) and expand \(\mathbf{G}_-\) perturbatively.

\begin{equation}
\mathbf{G}=\frac{1}{-i\omega+\epsilon_0(k)+2t_\perp}\left(1 + \frac{J \boldsymbol{\mathcal{M}}_+}{-i\omega+\epsilon_0(k)+2t_\perp}\right)^{-1} \approx \frac{1}{2t_\perp},
\end{equation}Inserting the expansion in Eq.(\ref{eq:strongintout}) gives for the Lagrangian of the lower band

\begin{equation}
\mathcal{L}= \ccd_+ \left[-\mathbf{G}_+^{-1}-\frac{J^2}{2 t_\perp} \left(\mathbf{S}^b_-\right)^2\right]\cc_+.
\end{equation}

\section{Suppl.Mat. B: Separating fast and slow variables}

In Eq.(\ref{eq:bsnu1nu2}), we obtained fast oscillating backscattering terms. However, we assume, that the spins \(\mathbf{S}_{1,2}\) have fast oscillating components, which can compensate the fastness of the backscattering. To see this, we explicitly separate the fast and the slow variables with a suitable parameterization for the spins. We start from Eq.(\ref{eq:bsnu1nu2})

\begin{align}
	\mathcal{L}_{bs}^{\nu\nu}&= \frac{J_a}{2} \Rd_\nu \mathbf{S}^a_+ \sigma_a \Ll_\nu e^{2i k_{F}^{(\nu)}x} ,\\
	\mathcal{L}_{bs}^{\nu\nu'}&= \frac{J_a}{2} \Rd_\nu \mathbf{S}^a_- \sigma_a \Ll_{\nu'} e^{i (k_{F}^{(+)}+k_{F}^{(-)})x}.
\end{align}  We explicitly single out a slow and fast component of the spin around \( \tilde{k}_F \equiv
( k_{F}^{(+)} + k_{F}^{(-)} ) /2\). This procedure does not result in over counting angles, since after integrating out the massive variables, the low energy theory only depends on two angles per spin, thus justifying this approach.

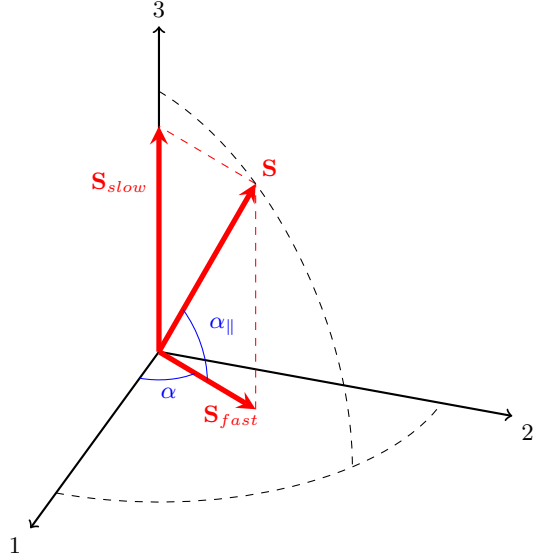
\begin{figure}[htbp]
	\input{figures/imspin}
	\caption{Decomposition of the individual impurity spins in a fast and slow component.}
	\label{fig:spinpara}
\end{figure}

\begin{equation}
\frac{\mathbf{S}^{(n)}}{s} = \left(\mathbf{m}_n + \left[\mathbf{e}_{1} \cos\left(2\tilde{k}_F x  \right)+\mathbf{e}_{2} \sin\left(2\tilde{k}_F x \right) \right] \sqrt{1-\mathbf{m}_n^2}\right)^{(n)}, \quad \mathbf{m}_n= \sin\left(\alpha^{(n)}\right) \mathbf{e}_{3}^{(n)}
\end{equation} Now we parameterize the orthonormal triad \(\{\mathbf{e}_{1}^{(n)},\mathbf{e}_{2}^{(n)},\mathbf{e}_{3}^{(n)}\}\) by spherical coordinates.

\begin{align}
	\mathbf{e}_{1}^{(n)}&=  \left(-\cos\left(\theta^{(n)}\right)\cos\left(\psi^{(n)}\right),-\cos\left(\theta^{(n)}\right)\sin\left(\psi^{(n)}\right),\sin\left(\theta^{(n)}\right)\right)^T \\
	\mathbf{e}_{2}^{(n)}&=  \left(\sin\left(\psi^{(n)}\right),-\cos\left(\psi^{(n)}\right),0\right)^T\\
	\mathbf{e}_{3}^{(n)}&=  \left(\sin\left(\theta^{(n)}\right)\cos\left(\psi^{(n)}\right),\sin\left(\theta^{(n)}\right)\sin\left(\psi^{(n)}\right),\cos\left(\theta^{(n)}\right)\right)^T
\end{align}.

After inserting the new spin parameterization, the back scattering terms take the following form

\begin{align}
\mathcal{L}_{bs}^{\nu\nu}&= \Rd_{\nu,\sigma} \left(\mathbf{\Delta}^{(1)}+\mathbf{\Delta}^{(2)}\right)  \Ll_{\nu,\sigma'} e^{i \nu \delta k_{F}x} ,\\
\mathcal{L}_{bs}^{\nu\nu'}&= \Rd_{\nu,\sigma} \left(\mathbf{\Delta}^{(1)}-\mathbf{\Delta}^{(2)}\right)  \Ll_{\nu',\sigma'}.
\end{align} The scattering amplitudes are given by
\begin{equation}
\mathbf{\Delta}^{(n)} = \tilde{J} \left[e^{i\psi^{(n)}}\sin^2\left(\frac{\theta^{(n)}}{2}\right)  \sigma_- - e^{-i\psi^{(n)}} \cos^2 \left(\frac{\theta^{(n)}}{2}\right) \sigma_+\right],
\end{equation} with \(\tilde{J}=s\rho_s \frac{J}{2} \cos \left(\alpha^{(n)}\right)\). The amplitudes contain the phase factors \(\psi^{(1)}\) and \(\psi^{(2)}\), which can be partially gauged away. The rest, especially \(\alpha^{(1)},\alpha^{(2)},\theta^{(1)},\theta^{(2)}\) enters the ground state energy equation. The classical values of the latter four angles are thus determined by the configuration which has the minimal ground state energy.

\label{ch:GSE1D}
\section{Suppl.Mat. C: Groundstate energy of gapped 1D Dirac fermions}
We want to calculate the gain in ground state energy of gapped 1D Dirac fermions with respect to the ungapped fermions. Let us consider a gapped fermionic Green's function of the form

\begin{equation} \label{eq:G}
-\mathbf{G}^{-1}=\begin{pmatrix}
-i \omega+v_{F_1} k & m \\
m^*&-i\omega-v_{F_2} k
\end{pmatrix},
\end{equation} and define

\begin{equation}
-\mathbf{G}_0^{-1}=\begin{pmatrix}
-i \omega+v_{F_1} k & 0 \\
0&-i\omega-v_{F_2} k
\end{pmatrix}, \quad \mathbf{\Delta}=\begin{pmatrix}
0 & m \\
m^*& 0
\end{pmatrix}.
\end{equation}

The partition function corresponding to Eq.(\ref{eq:G}) is given by

\begin{equation}
Z=\det\left(-\mathbf{G}^{-1}\right)=\det\left(-\mathbf{G}_0^{-1}+\mathbf{\Delta}\right)=Z_0 \exp\operatorname{Tr}\log\left(1-\mathbf{G}_0\mathbf{\Delta}\right),
\end{equation} where we used the identity \(\det\left(\mathbf{A}\right)=\exp\operatorname{Tr}\log\left(\mathbf{A}\right)\) in the last step. We can compute the free energy \(F=-T\log\left(Z\right)\), and expand the free energy in leading order of \(\Delta\). We find

\begin{equation}\label{eq:gSc}
F=F_0-T\operatorname{Tr}\log\left(1-\mathbf{G}_0\mathbf{\Delta}\right) \approx F_0 + \frac{T}{2} \operatorname{Tr} \mathbf{G}_0\mathbf{\Delta}\mathbf{G}_0\mathbf{\Delta}.
\end{equation} Note that the linear term in the expansion is absent, because of the off diagonal structure of \(\mathbf{\Delta}\) and on the other hand reflects the fact, that we expand the ground state energy around its minimum. In the limit \(T\rightarrow 0\) we can convert the summation over the Matsubara frequency to an integral and find

\begin{equation}
\delta E = 	\frac{T}{2} \operatorname{Tr} \mathbf{G}_0\mathbf{\Delta}\mathbf{G}_0\mathbf{\Delta} \rightarrow - \xi_0 \int \frac{d\{\omega,k\}}{(2\pi)^2} \frac{|m|^2}{(-i\omega+v_{F_1}k)(-i\omega-v_{F_2}k)},
\end{equation} which has poles at \(\omega_1=- i v_{F_{1}} k\) and  \(\omega_2= i v_{F_{2}} k\). We find

\begin{equation}\label{eq:Gse}
\delta E = -\frac{\xi_0}{2\pi} \int^{2t}_{|m|} \mathop{dk} \frac{|m|^2}{k\left(v_{F_1}+v_{F_2}\right)} = -\frac{\xi_0}{2\pi\left(v_{F_1}+v_{F_2}\right)} |m|^2\log\left(\frac{2t}{|m|}\right),
\end{equation} where we used the band width as a high energy cut-off.

%
%
%
%
%
%
%

\section{Suppl.Mat. D: Derivation of the ground state energy equation in the intermediate tunneling regime}

Let us consider the gap structure  \(\hat{m}_-\) of Eqs. (\ref{eq:helgap1}) and (\ref{eq:helgap2}). Our goal is to calculate and compare the ground state energies for both spin configurations. The gap structure \(\hat{m}_-\) plays the role of a mass term which mixes fermions of the \(\nu=\pm\) bands \(\mathcal{L}_{bs}^{\nu\nu'} \simeq
\Rd_{\nu} \hat{m}_- \Ll_{\nu'} + h.c., \ \nu \neq \nu'\).

\begin{eqnarray}\label{eq:helgap11}
\theta^{(1)} = \theta^{(2)} = 0: & & \hat{m}_- = 2i \tilde{J} e^{-i \tilde{\psi}} \! \sin\left(\delta\psi\right)\sigma_+ , \\
\theta^{(1)} = 0, \, \theta^{(2)} = \pi: & \ & \hat{m}_- = - \tilde{J} \left(e^{i\psi^{(2)}} \! \sigma_- + e^{-i\psi^{(1)}} \! \sigma_+ \right).\label{eq:helgap12}
\end{eqnarray}  The common phase factors \(\tilde{\psi}\) and  \(\psi^{(1)}, \psi^{(2)}\) in (\ref{eq:helgap11}) and (\ref{eq:helgap12}) can be removed by a gauge transformation or by bosonizing the theory and shifting the phases. The phases enter the low energy Lagrangian as a chiral anomaly in the form of a Luttinger liquid Lagrangian  \(\mathcal{L}_{\rm LL}[\Phi,v_\Phi] = [ (\partial_\tau \Phi)^2 + (v_\Phi \partial_x \Phi)^2 ] / 2 \pi K_\Phi, \ \Phi =\tilde{\psi},\psi^{(1)}, \psi^{(2)} \), which we will discuss later. We set the common phase factors to zero in the following calculations.\newline

{\noindent\bf Spin configuration  I: \(\theta^{(1)} = \theta^{(2)} = 0\)}\newline

The gap structure gaps only fermions of helicity \(h=+1\) and is given by

\begin{equation}
	\hat{m}_- = m_- \sigma_+, \quad m_- = 2i \tilde{J} \sin\left(\delta\psi\right), \label{eq:m-1}
\end{equation} which leads to the following inverse Green's function.

\begin{equation}
- \mathbf{G}^{-1}=\begin{array}{c cccc |  cccc  c c   }
	\ldelim({8}{1mm}&\partial_{R_+} & 0 & 0 & m_- & 0 & 0 & 0 & 0 &\hspace{-.5cm} \rdelim){8}{-5mm} & R_{+\uparrow} \\
	&0 & \partial_{L_+} & m^*_- & 0 & 0  & 0& 0 & 0 & & L_{+\downarrow} \\
	&0 & m_- & \partial_{R_-} &0  & 0 & 0 & 0 & 0 & &R_{-\uparrow} \\
	&m^*_- & 0 &0  & \partial_{L_-} & 0 & 0 & 0 & 0 && L_{-\downarrow} \\ \cline{2-9}&&&&&&&&&\\[-10pt]
	&0 & 0 & 0 & 0 & \partial_{R_+} & 0  & 0 &  0 & &R_{+\downarrow} \\
	&0 & 0 & 0 & 0 & 0 & \partial_{L_+} & 0 & 0 & &L_{+\uparrow} \\
	&0 & 0 & 0 & 0 & 0 & 0 & \partial_{R_-} & 0 & &R_{-\downarrow} \\
	&0 & 0 & 0 & 0 & 0 &0 & 0  & \partial_{L_-}  & &L_{-\uparrow}
	\end{array},
\vspace{.5cm}
\end{equation} where \(\partial_{\nicefrac{R}{L}\nu} = \partial_\tau \mp i v_{F_\nu}\partial_x \) is the chiral derivative for the respective bands and the ordering of the states is indicated to the right of the Green's function. We focus on the gapped block

\begin{equation}\label{eq:GFint1}
- \mathbf{G}_m^{-1}=\begin{array}{c cccc   c c   }
\ldelim({4}{1mm}&\partial_{R_+} & 0 & 0 & m_-  &\hspace{-.5cm} \rdelim){4}{-5mm} & R_{+\uparrow} \\
&0 & \partial_{L_+} & m^*_- & 0   & & L_{+\downarrow} \\
&0 & m_- & \partial_{R_-} &0  &  &R_{-\uparrow} \\
&m^*_- & 0 &0  & \partial_{L_-}   & & L_{-\downarrow} \\
\end{array}, \overset{\mathbf{U}}{\rightarrow}  \begin{array}{c cccc   c c   }
\ldelim({4}{1mm}&\partial_{R_+} & m_- & 0 & 0  &\hspace{-.5cm} \rdelim){4}{-5mm} & R_{+\uparrow} \\
&m^*_- & \partial_{L_-} & 0 & 0   & & L_{-\downarrow} \\
&0 & 0 & \partial_{R_-} &m^*_-  &  &R_{-\uparrow} \\
&0 & 0 &m_-  & \partial_{L_+}   & & L_{+\downarrow} \\
\end{array}
\end{equation}

Changing the ordering of the states gives us the above matrix in block diagonal form. This allows us to use Eq.(\ref{eq:Gse}), see Suppl.Mat.C. We now integrate out all gapped fermions. The ground state energy is the sum of the ground state energy of the two individual blocks in (\ref{eq:GFint1}), which gives a factor of two compared to (\ref{eq:Gse}).

\begin{equation}\label{eq:GSitI}
\delta E^{\rm (IT,I)} = - \frac{4 \xi_0 \tilde{J}^2 \sin^2\left(\delta\psi\right)}{\pi \left(v_{F_+}+v_{F_-}\right)}\log\left(\frac{2t}{|\tilde{J}|}\right);
\end{equation} The ground state energy is minimal if there is a mode-locking of the in-plane polar angles \(\delta\psi=\pm \frac{\pi}{2}\).\newline

{\noindent\bf Spin configuration  II: \(\theta^{(1)} = 0, \, \theta^{(2)} = \pi\)}\newline

The gap structure now contains gaps for fermions of all helicites in both bands, but the effective size of the gap is reduced by a factor of two.

\begin{equation}
\hat{m}_- = m_- \left(\sigma_- + \sigma_+ \right), \quad m_-=-\tilde{J}.
\end{equation} This leads to the following inverse Green's function.

\begin{equation}
- \mathbf{G}^{-1}=\begin{array}{c cccc |  cccc  c c   }
\ldelim({8}{1mm}&\partial_{R_+} & 0 & 0 & m_- & 0 & 0 & 0 & 0 &\hspace{-.5cm} \rdelim){8}{-5mm} & R_{+\uparrow} \\
&0 & \partial_{L_+} & m_- & 0 & 0  & 0& 0 & 0 & & L_{+\downarrow} \\
&0 & m_- & \partial_{R_-} &0  & 0 & 0 & 0 & 0 & &R_{-\uparrow} \\
&m_- & 0 &0  & \partial_{L_-} & 0 & 0 & 0 & 0 && L_{-\downarrow} \\ \cline{2-9}&&&&&&&&&\\[-10pt]
&0 & 0 & 0 & 0 & \partial_{R_+} & 0  & 0 &   m_- & &R_{+\downarrow} \\
&0 & 0 & 0 & 0 & 0 & \partial_{L_+} &  m_- & 0 & &L_{+\uparrow} \\
&0 & 0 & 0 & 0 & 0 &  m_- & \partial_{R_-} & 0 & &R_{-\downarrow} \\
&0 & 0 & 0 & 0 &  m_- & 0 & 0  & \partial_{L_-}  & &L_{-\uparrow}
\end{array},
\vspace{.5cm}
\end{equation}

It can be block diagonalized in the following form

\begin{equation}
- \mathbf{G}^{-1}=\begin{array}{c cccc |  cccc  c c   }
\ldelim({8}{1mm}&\partial_{R_+} & m_- & 0 & 0 & 0 & 0 & 0 & 0 &\hspace{-.5cm} \rdelim){8}{-5mm} & R_{+\uparrow} \\
&m_- & \partial_{L_-} & 0 & 0 & 0  & 0& 0 & 0 & & L_{-\downarrow} \\
&0 & 0 & \partial_{R_-} &m_-  & 0 & 0 & 0 & 0 & &R_{-\uparrow} \\
&0 & 0 &m_-  & \partial_{L_+} & 0 & 0 & 0 & 0 && L_{+\downarrow} \\ \cline{2-9}&&&&&&&&&\\[-10pt]
&0 & 0 & 0 & 0 & \partial_{R_+} & m_-  & 0 & 0 & &R_{+\downarrow} \\
&0 & 0 & 0 & 0 & m_- & \partial_{L_-} &  0 & 0 & &L_{-\uparrow} \\
&0 & 0 & 0 & 0 & 0 &  0 & \partial_{R_-} & m_- & &R_{-\downarrow} \\
&0 & 0 & 0 & 0 &  0 & 0 & m_-  & \partial_{L_+}  & &L_{+\uparrow}
\end{array}.
\vspace{.5cm}
\end{equation} We again integrate out all gapped fermions with the help of Eq. (\ref{eq:Gse}) and obtain
\begin{equation}\label{eq:GSitII}
\delta E^{\rm (IT,II)} = - \frac{2 \xi_0 \tilde{J}^2}{\pi \left(v_{F_+}+v_{F_-}\right)}\log\left(\frac{2t}{|\tilde{J}|}\right);
\end{equation}

Comparing (\ref{eq:GSitI}) and (\ref{eq:GSitII}), we find, that a helical phase, where fermions with helicity \(h=-1\) remain gapless, is energetically favored \(\delta E^{\rm (IT,I)} < \delta E^{\rm (IT,II)}\). The case where the gapless modes have helicity \(h=+1\) can be found analogously with \(\theta^{(1)} = \theta^{(2)} = \pi\)

\section{Suppl.Mat. E: Derivation of the ground state energy equation in the weak tunneling regime}

Let us now consider the weak tunneling regime. In contrast to the intermediate tunneling case we find the "off diagonal" gap structure \(\mathcal{L}_{bs}^{\nu\nu'} \simeq
\Rd_{\nu} \hat{m}_- \Ll_{\nu'} + h.c., \ \nu \neq \nu'\) of Eqs. (\ref{eq:helgap1}) and (\ref{eq:helgap2}) and the "diagonal" gaps \(\mathcal{L}_{bs}^{\nu\nu} \simeq
\Rd_{\nu} \hat{m}_+ \Ll_{\nu} + h.c.\) of Eqs. (\ref{eq:helgap1+}) and (\ref{eq:helgap2+}). This makes the weak tunneling case distinct from the intermediate tunneling case. Similar to Suppl.Mat. D, we analyze two different spin configuration and calculate and compare the ground state energies for both spin configurations.

\begin{align}\label{eq:helgap21}
\theta^{(1)} = \theta^{(2)} = 0 &: \hat{m}_- = 2i \tilde{J} e^{-i \tilde{\psi}}  \sin\left(\delta\psi\right)\sigma_+ , \\
\theta^{(1)} = \theta^{(2)} = 0 &: \hat{m}_+ = -2 \tilde{J} e^{-i \tilde{\psi}} \cos\left(\delta\psi\right)\sigma_+ ;\label{eq:helgap11+}
\end{align}

\begin{align}
\theta^{(1)} = 0, \, \theta^{(2)} = \pi &: \hat{m}_- = - \tilde{J} \left(e^{i\psi^{(2)}}  \sigma_- + e^{-i\psi^{(1)}} \! \sigma_+ \right).\label{eq:helgap22}       \\
\label{eq:helgap12+}
\theta^{(1)} = 0, \, \theta^{(2)} = \pi &: \hat{m}_+ = \tilde{J} \left(e^{i\psi^{(2)}} \sigma_--e^{-i\psi^{(1)}} \sigma_+\right);
\end{align}

The common phase factors \(\tilde{\psi}\) and  \(\psi^{(1)}, \psi^{(2)}\) in Eqs.(\ref{eq:helgap21}) - (\ref{eq:helgap12+}) can be removed by a gauge transformation or by bosonizing the theory and shifting the phases. The phases enter the low energy Lagrangian in the form of a Luttinger liquid Lagrangian  \(\mathcal{L}_{\rm LL}[\Phi,v_\Phi] = [ (\partial_\tau \Phi)^2 + (v_\Phi \partial_x \Phi)^2 ] / 2 \pi K_\Phi, \ \Phi =\tilde{\psi},\psi^{(1)}, \psi^{(2)} \), which we will discuss later. We set the common phase factors to zero in the following calculations.\newline

{\noindent\bf Spin configuration  I: \(\theta^{(1)} = \theta^{(2)} = 0\)}\newline

The gap structure gaps only fermions of helicity \(h=+1\) and is given by

\begin{align}
\hat{m}_- = m_- \sigma_+ &, \quad m_- = 2i \tilde{J} \sin\left(\delta\psi\right), \label{eq:m1}\\
\hat{m}_+ = m_+ \sigma_+ &, \quad m_+ = -2 \tilde{J} \cos\left(\delta\psi\right),  \label{eq:m2}
\end{align} which leads to the following inverse Green's function.

\begin{equation}
- \mathbf{G}^{-1}=\begin{array}{c cccc |  cccc  c c   }
\ldelim({8}{1mm}&\partial_{R_+} & m_+ & 0 & m_- & 0 & 0 & 0 & 0 &\hspace{-.5cm} \rdelim){8}{-5mm} & R_{+\uparrow} \\
&m_+ & \partial_{L_+} & m^*_- & 0 & 0  & 0& 0 & 0 & & L_{+\downarrow} \\
&0 & m_- & \partial_{R_-} &m_+  & 0 & 0 & 0 & 0 & &R_{-\uparrow} \\
&m^*_- & 0 &m_+  & \partial_{L_-} & 0 & 0 & 0 & 0 && L_{-\downarrow} \\ \cline{2-9}&&&&&&&&&\\[-10pt]
&0 & 0 & 0 & 0 & \partial_{R_+} & 0  & 0 &  0 & &R_{+\downarrow} \\
&0 & 0 & 0 & 0 & 0 & \partial_{L_+} & 0 & 0 & &L_{+\uparrow} \\
&0 & 0 & 0 & 0 & 0 & 0 & \partial_{R_-} & 0 & &R_{-\downarrow} \\
&0 & 0 & 0 & 0 & 0 &0 & 0  & \partial_{L_-}  & &L_{-\uparrow}
\end{array},
\vspace{.5cm}
\end{equation} where \(\partial_{\nicefrac{R}{L}\nu} = \partial_\tau \mp i v_{F_\nu}\partial_x \) is the chiral derivative for the respective bands and the ordering of the states is indicated to the right of the Green's function. We focus on the gapped block

\begin{equation}\label{eq:GFint11}
- \mathbf{G}_m^{-1}=\begin{array}{c cccc   c c   }
\ldelim({4}{1mm}&\partial_{R_+} & m_+ & 0 & m_-  &\hspace{-.5cm} \rdelim){4}{-5mm} & R_{+\uparrow} \\
&m_+ & \partial_{L_+} & m^*_- & 0   & & L_{+\downarrow} \\
&0 & m_- & \partial_{R_-} &m_+  &  &R_{-\uparrow} \\
&m^*_- & 0 &m_+  & \partial_{L_-}   & & L_{-\downarrow} \\
\end{array}.
\end{equation} In contrast to the intermediate tunneling case, we cannot block diagonalize this matrix and make use of the Eq. (\ref{eq:Gse}) in Suppl.Mat. C. However, we can directly integrate out the gapped fermions analogously to Eq.(\ref{eq:gSc}).

\begin{gather}\label{eq:m+-}
	\delta E^{(WT,I)} \simeq \frac{T}{2} \operatorname{Tr}\mathbf{G}_0\mathbf{\Delta}\mathbf{G}_0\mathbf{\Delta} = \frac{T}{2} \operatorname{Tr}\mathbf{G}_0\mathbf{\Delta}_+\mathbf{G}_0\mathbf{\Delta}_+ +\frac{T}{2} \operatorname{Tr}\mathbf{G}_0\mathbf{\Delta}_-\mathbf{G}_0\mathbf{\Delta}_-;\\
	\mathbf{G}_0 = \begin{pmatrix}
	\partial_{R_+} & 0& 0 & 0   \\
	0 & \partial_{L_+} & 0 & 0  \\
	0 & 0 & \partial_{R_-} &0   \\
	0 & 0 &0  & \partial_{L_-}
	\end{pmatrix}^{-1}, \ \mathbf{\Delta}= \underbrace{\begin{pmatrix}
	0 & m_+ & 0 & 0  \\
	m_+ & 0 & 0 & 0   \\
	0 & 0& 0 &m_+  \\
	0 & 0 &m_+  & 0
	\end{pmatrix}}_{\mathbf{\Delta}_+}+\underbrace{\begin{pmatrix}
	0 & 0 & 0 & m_-  \\
	0 & 0 & m^*_- & 0   \\
	0 & m_- & 0 & 0  \\
	m^*_- & 0 & 0  & 0
	\end{pmatrix}}_{\mathbf{\Delta}_-}
\end{gather} Note, that the mixing terms \(\operatorname{Tr}\mathbf{G}_0\mathbf{\Delta}_\pm\mathbf{G}_0\mathbf{\Delta}_\mp\) in Eq.(\ref{eq:m+-}) vanish, due to the off diagonal structure of the matrix \(\mathbf{\Delta}_-\). The ground state energy is given as the sum of the ground state energies coming from the respective gap structures \(\hat{m}_\pm\), which can both be computed now using Eq. (\ref{eq:Gse}) in Suppl.Mat. C. The addition to the ground state energy coming from the off diagonal matrix \(\mathbf{\Delta}_-\) has already be considered in Suppl.Mat. D. We integrate out all gapped fermions and find

\begin{multline}\label{eq:GSWTI}
	\delta E^{(WT,I)}=   -\frac{\xi_0 \tilde{J}^2}{\pi}\left[\underbrace{\frac{ \cos^2\left(\delta\psi\right)}{v_{F_+}}+\frac{ \cos^2\left(\delta\psi\right)}{v_{F_-}}}_{\operatorname{Tr}\mathbf{G}_0\mathbf{\Delta}_+\mathbf{G}_0\mathbf{\Delta}_+}+\underbrace{\frac{4 \sin^2\left(\delta\psi\right)}{\left(v_{F_+}+v_{F_-}\right)}}_{\operatorname{Tr}\mathbf{G}_0\mathbf{\Delta}_-\mathbf{G}_0\mathbf{\Delta}_-}\right]\log\left(\frac{2t}{|\tilde{J}|}\right)=\\=
	-\frac{\xi_0 \tilde{J}^2}{2\pi}\left[\frac{(v_{F_+}+v_{F_-})^2 +(v_{F_+}-v_{F_-})^2\cos\left(2\delta\psi\right)+4v_{F_+}v_{F_-} }{v_{F_+}v_{F_-}(v_{F_+}+v_{F_-})}\right]\log\left(\frac{2t}{|\tilde{J}|}\right).
\end{multline}

The ground state energy is minimal if there is a mode-locking of the in-plane polar angles of \(\delta\psi=0,\pi \). In these cases intraband scattering processes dominate and interband scattering vanishes. However, Eq.(\ref{eq:GSWTI}) contains terms which are beyond the accuracy set by the scale separation. We use the smallness of \( \xi_0 t_\perp / \xi_{EP}\tilde{J} \ll 1 \), and expand our result perturbatively, which gives

\begin{equation}
	\label{EwtI}
	\delta E^{\rm (WT,I)} = -\frac{2\xi_0 \tilde{J}^2}{\pi \tilde{v}_F}
	\left[1 + 4t_\perp^2\left(\frac{\xi_0^2}{\tilde{v}_F^3}+\frac{\left(1+\cos\left(2\delta\psi\right)\right)\xi_0^4\tilde{\varepsilon}_F^2}{\tilde{v}_F^5}\right)\right]
	\log\left(\frac{2t}{|\tilde{J}|}\right),
\end{equation} where \(\tilde{v}_F=\left.v_{F_\pm}\right|_{t_\perp=0}\) and \(\tilde{\varepsilon}_F=\varepsilon(k)\). \newline

{\noindent\bf Spin configuration  II: \(\theta^{(1)} = 0, \, \theta^{(2)} = \pi\)}\newline

The gap structure now contains gaps for fermions of all helicites in both bands, but the effective size of the gap is reduced by a factor of two.

\begin{align}
\hat{m}_- = m_- \left(\sigma_- + \sigma_+ \right) &, \quad m_-=-\tilde{J},\\
\hat{m}_+ = m_+ \left(\sigma_- - \sigma_+\right)  &, \quad m_+= \tilde{J},
\end{align} which leads to the following inverse Green's function.

\begin{equation}
- \mathbf{G}^{-1}=\begin{array}{c cccc |  cccc  c c   }
\ldelim({8}{1mm}&\partial_{R_+} & -m_+ & 0 & m_- & 0 & 0 & 0 & 0 &\hspace{-.5cm} \rdelim){8}{-5mm} & R_{+\uparrow} \\
&-m_+ & \partial_{L_+} & m_- & 0 & 0  & 0& 0 & 0 & & L_{+\downarrow} \\
&0 & m_- & \partial_{R_-} &-m_+  & 0 & 0 & 0 & 0 & &R_{-\uparrow} \\
&m_- & 0 &-m_+  & \partial_{L_-} & 0 & 0 & 0 & 0 && L_{-\downarrow} \\ \cline{2-9}&&&&&&&&&\\[-10pt]
&0 & 0 & 0 & 0 & \partial_{R_+} & m_+  & 0 &  m_- & &R_{+\downarrow} \\
&0 & 0 & 0 & 0 & m_+ & \partial_{L_+} & m_- & 0 & &L_{+\uparrow} \\
&0 & 0 & 0 & 0 & 0 & m_- & \partial_{R_-} & m_+ & &R_{-\downarrow} \\
&0 & 0 & 0 & 0 & m_- & 0 & m_+  & \partial_{L_-}  & &L_{-\uparrow}
\end{array},
\vspace{.5cm}
\end{equation}

Similar to the other spin configuration, we integrate out all gapped fermions and find

\begin{multline}
	\delta E^{(WT,II)}= -\frac{\xi_0 \tilde{J}^2}{4\pi}\left[\underbrace{\frac{2}{v_{F_+}}+\frac{2}{v_{F_-}}}_{\operatorname{Tr}\mathbf{G}_0\mathbf{\Delta}_+\mathbf{G}_0\mathbf{\Delta}_+}+\underbrace{\frac{8}{\left(v_{F_+}+v_{F_-}\right)}}_{\operatorname{Tr}\mathbf{G}_0\mathbf{\Delta}_-\mathbf{G}_0\mathbf{\Delta}_-}\right]\log\left(\frac{2t}{|\tilde{J}|}\right)=\\=-\frac{\xi_0 \tilde{J}^2}{2\pi}\left[\frac{(v_{F_+}+v_{F_-})^2 +4v_{F_+}v_{F_-} }{v_{F_+}v_{F_-}(v_{F_+}+v_{F_-})}\right]\log\left(\frac{2t}{|\tilde{J}|}\right).
\end{multline} We again expand perturbatively and obtain

\begin{equation}
\label{EwtII}
\delta E^{\rm (WT,II)} = -\frac{2\xi_0 \tilde{J}^2}{\pi \tilde{v}_F}
\left[1 + 4t_\perp^2\left(\frac{\xi_0^2}{\tilde{v}_F^3}+\frac{\xi_0^4\tilde{\varepsilon}_F^2}{\tilde{v}_F^5}\right)\right]
\log\left(\frac{2t}{|\tilde{J}|}\right),.
\end{equation}

Similar to the intermediate tunneling case, the mode locking of the in-plane polar angle \(\delta\psi=0\) provides the minimum of the ground state energy in the weak tunneling case \( t_\perp \leq J \ll 2t\) and makes spin configuration I more favorable. In the limit of vanishing tunneling \( t_\perp \leq J \ll 2t\), we neglect higher order terms and Spin configuration I and II cannot be distinguished. In Suppl.Mat. G we introduce an additional Dresselhaus SOI and show that it removes the ground state degeneracy in favor of the helical phase. The wires behave essentially uncoupled. Spin configuration I resembles a helical wire where all channels support fermions of the same helicity, either \(h=+1\) or \(h=-1\). In Spin configuration II, fermions can travel in any direction. Some channels have helicity \(h=+1\) and others \(h=-1\). This configuration resembles a non-helical spinful wire.

\section{Suppl.Mat. F: Spinless disorder}

Let us consider the influence of disorder, modeled by a weak random scalar potential in each wire. On the Hamiltonian level we get the interaction term

\begin{equation}
	\hat{H}_{\rm dis} = \int dx \  V(x) \left(\ccd(x)\cc(x) \right)^{(n)} = \int dk \int dq \ V(q)\left( \ccd(k+q) \cc(k) \right)^{(n)},
\end{equation}where we defined \(V_q=\int dx\ e^{iqx}\ V(x)\). We switch from the wire index to the band index using the transformation, which diagonalized the non-interacting Hamiltonian \(\cc_{\nicefrac{1}{2}}=\frac{1}{\sqrt{2}}\left(\cc_+\pm \cc_-\right)\) and obtain

\begin{equation}
\hat{H}_{\rm dis} =  \int dk \int dq \ V(q)\left( \ccd(k+q)_+ \cc(k)_+ + \ccd(k+q)_- \cc(k)_- \right),
\end{equation} Since we are interested in the low energy behavior of the system we focus on the momenta around the Fermi points of the system \(q=\pm 2k_{F_\pm}\). This gives

\begin{equation}
\hat{H}_{\rm dis} =  \int dk  \left( \sum_{q=\pm 2k_{F_+}} \ V(q)\ \ccd(k+q)_+ \cc(k)_+ + \sum_{q=\pm 2k_{F_-}} \ V(q)\ \ccd(k+q)_- \cc(k)_- \right),
\end{equation}
We now switch to the Lagrangian formulation and introduce the smooth chiral fields \(\Rr_\pm(k)=\psi_{k+k_{F_\pm}}\) and \(\Ll_\pm(k)=\psi_{k-k_{F_\pm}}\), which are the  shifted fermionic fields \(\psi_\pm\) of the disorder Lagrangian. We obtain

\begin{equation}\label{eq:Ldis}
	\mathcal{L}_{\rm dis} =  g_{2k_{F_+}}\Rd_{+} \Ll_{+} + g_{2k_{F_-}} \Rd_{-} \Ll_{-} +h.c.,
\end{equation} where \(g_{2k_{F_\pm}}=V(\pm 2k_{F_\pm})\). We analyze the case where we find four Fermi points on the level of the chemical potential. It is convenient to analyze the intermediate and weak tunneling regimes individually.\newline

{\noindent \bf Weak tunneling regime}\newline

For simplicity, we neglect the difference between the Fermi momenta \(2k_{F_+}\approx 2k_{F_-}\), since the Fermi points almost coincide in pairs. Furthermore, we assume that the system adapts the spin configuration I: \(\theta^{(1)} = \theta^{(2)} = 0\). The Green's function of the system is given by

\begin{equation}
- \mathbf{G}^{-1}=\begin{array}{c cccc |  cccc  c c   }
\ldelim({8}{1mm}&\partial_{R_+} & m_+ & 0 & m_- & 0 & g_{2k_{F_+}} & 0 & 0 &\hspace{-.5cm} \rdelim){8}{-5mm} & R_{+\uparrow} \\
&m_+ & \partial_{L_+} & m^*_- & 0 & g_{2k_{F_+}}  & 0& 0 & 0 & & L_{+\downarrow} \\
&0 & m_- & \partial_{R_-} &m_+  & 0 & 0 & 0 & g_{2k_{F_-}} & &R_{-\uparrow} \\
&m^*_- & 0 &m_+  & \partial_{L_-} & 0 & 0 & g_{2k_{F_-}} & 0 && L_{-\downarrow} \\ \cline{2-9}&&&&&&&&&\\[-10pt]
&0 & g_{2k_{F_+}} & 0 & 0 & \partial_{R_+} & 0  & 0 &  0 & &R_{+\downarrow} \\
&g_{2k_{F_+}} & 0 & 0 & 0 & 0 & \partial_{L_+} & 0 & 0 & &L_{+\uparrow} \\
&0 & 0 & 0 & g_{2k_{F_-}} & 0 & 0 & \partial_{R_-} & 0 & &R_{-\downarrow} \\
&0 & 0 & g_{2k_{F_-}} & 0 & 0 &0 & 0  & \partial_{L_-}  & &L_{-\uparrow}
\end{array},
\vspace{.5cm}
\end{equation} with \( m_+ = -2 \tilde{J} \cos\left(\delta\psi\right), \ m_- = 2i \tilde{J} \sin\left(\delta\psi\right) \). Our goal is to integrate our the gapped fermions and derive an effective action for the ungapped fermions.  We define the gapped spinor \(\Psi_g=\left( R_{+\uparrow},L_{+\downarrow}, R_{-\uparrow},L_{-\downarrow}\right)^T\) and the ungapped spinors \(\Psi_u=\left( R_{+\downarrow},L_{+\uparrow}, R_{-\downarrow},L_{-\uparrow}\right)^T\). Straightforward calculation yields

\begin{equation}\label{eq:disgap}
	\left\langle \exp \left\{-\int d\{\tau,x\}  \Psi^\dagger_g \ \hat{g} \ \Psi^{\vphantom{\dagger}}_u\right\} \right\rangle_g=\exp \left\{\frac{1}{2}\int d\{\tau_1,x_1,\tau_2,x_2\}  \left(\hat{g}\Psi_u\right)_{(x_1,\tau_1)}^{\dagger} \left\langle \Psi_g^\dagger\Psi_g^{\vphantom{\dagger}}\right\rangle_{g \ \left(\substack{x_1-x_2\\\tau_1-\tau_2}\right)} \left(\hat{g}\Psi_u\right)_{(x_2,\tau_2)}\right\}.
\end{equation} \(\langle\dots\rangle_g\) denotes the integration over the gapped fermions and we defined \(\hat{g}=\left(g_{2k_{F_+}}\sigma_+\sigma_-+g_{2k_{F_-}}\sigma_-\sigma_+\right)\otimes \tau_x\), where \(\sigma_\pm = \sigma_{x}\pm i \sigma_{y}\) with \(\sigma_{x,y},\tau_{x,y}\) being the first and second pauli matrix in spin and wire band space. The gapped Green's function can be computed directly by inversion if we insert the results we got earlier, namely, the in-plane modelocking of the form \(\delta\psi=0\) and \(v_{F_+}=v_{F_-}=v_F\). Note, that after the modelocking we only find intraband scattering \(m_+=2\tilde{J}\) and \(m_-=0\). This gives

\begin{equation}
	\left\langle \Psi_g^\dagger\Psi_g^{\vphantom{\dagger}}\right\rangle_g = \mathbf{G}_m=\frac{1}{4 \tilde{J}^2+v_F^2 k^2 +\omega^2} \begin{pmatrix} i\omega + v_F k & 2\tilde{J} & 0 & 0\\
	2 \tilde{J} & i\omega - v_F k & 0 & 0\\
	0&0& i\omega + v_F k & 2\tilde{J} \\
	0&0& 2\tilde{J} & i\omega - v_F k
	\end{pmatrix}\approx \begin{pmatrix} 0 & \frac{1}{2\tilde{J}} & 0 & 0\\
	\frac{1}{2\tilde{J}} & 0 & 0 & 0\\
	0&0& 0 & \frac{1}{2\tilde{J}} \\
	0&0& \frac{1}{2\tilde{J}} & 0
	\end{pmatrix},
\end{equation} which we expanded using the smallness of \(\frac{t}{\tilde{J}}\). Using Eq.(\ref{eq:disgap}) gives for the effective action of the ungapped fermions

\begin{equation}
	\mathcal{L}_{\rm dis} \simeq \frac{g_{2k_{F_+}}^2 \! (x)}{4\tilde{J}}\Rd_{+\downarrow}\Ll_{+\uparrow}+\frac{g_{2k_{F_-}}^2 \! (x)}{4\tilde{J}}\Rd_{-\downarrow}\Ll_{-\uparrow}+h.c.
\end{equation}If \(g_{2k_{F_\pm}}/\tilde{J}\ll1\), effective backscattering, which is governed by multiparticle scattering processes and localization  is suppressed. The transport properties of the gapless modes become protected up to parametrically large sample sizes. In spin configuration II \(\theta^{(1)} = 0, \, \theta^{(2)} = \pi\) all fermionic modes are gapped. In this case, there is no ballistic transport.\newline

{\noindent\bf Intermediate tunneling regime}\newline

In the intermediate tunneling regime, there are four Fermi points, which are well separated, i.e. the difference between the Fermi momenta is large \(\delta k_F \simeq \tilde{k}_F\). Let us start from Eq.(\ref{eq:Ldis})

\begin{equation}
\mathcal{L}_{\rm dis} =  g_{2k_{F_+}}\Rd_{+} \Ll_{+} + g_{2k_{F_-}} \Rd_{-} \Ll_{-} +h.c.
\end{equation}

Furthermore, we assume, that the system is in the energetically more favorable spin configuration I \(\theta^{(1)} = \theta^{(2)} = 0\), with a mode locking of the form \(\delta\psi=\pm\frac{\pi}{2}\). The Green's function is then given by

\begin{equation}
- \mathbf{G}^{-1}=\begin{array}{c cccc |  cccc  c c   }
\ldelim({8}{1mm}&\partial_{R_+} & 0 & 0 & m_- & 0 & g_{2k_{F_+}} & 0 & 0 &\hspace{-.5cm} \rdelim){8}{-5mm} & R_{+\uparrow} \\
&0 & \partial_{L_+} & m^*_- & 0 & g_{2k_{F_+}}  & 0& 0 & 0 & & L_{+\downarrow} \\
&0 & m_- & \partial_{R_-} &0  & 0 & 0 & 0 & g_{2k_{F_-}} & &R_{-\uparrow} \\
&m^*_- & 0 &0  & \partial_{L_-} & 0 & 0 & g_{2k_{F_-}} & 0 && L_{-\downarrow} \\ \cline{2-9}&&&&&&&&&\\[-10pt]
&0 & g_{2k_{F_+}} & 0 & 0 & \partial_{R_+} & 0  & 0 &  0 & &R_{+\downarrow} \\
&g_{2k_{F_+}} & 0 & 0 & 0 & 0 & \partial_{L_+} & 0 & 0 & &L_{+\uparrow} \\
&0 & 0 & 0 & g_{2k_{F_-}} & 0 & 0 & \partial_{R_-} & 0 & &R_{-\downarrow} \\
&0 & 0 & g_{2k_{F_-}} & 0 & 0 &0 & 0  & \partial_{L_-}  & &L_{-\uparrow}
\end{array},
\vspace{.5cm}
\end{equation} We proceed by integrating out the gapped fermions similar to the intermediate tunneling case using Eq.(\ref{eq:disgap}). The gapped Green's function is given by

\begin{equation}
\left\langle \Psi_g^\dagger\Psi_g^{\vphantom{\dagger}}\right\rangle_g=\begin{pmatrix} \frac{i\omega + v_{F_-} k}{4\tilde{J}^2+\omega^2 + i \omega k (v_{F_+}-v_{F_-})} &0 & 0 & \frac{ 2i\tilde{J}}{4\tilde{J}^2+\omega^2 + i \omega k (v_{F_+}-v_{F_-})}\\
0 & \frac{i\omega - v_{F_-} k}{4\tilde{J}^2+\omega^2 - i \omega k (v_{F_+}-v_{F_-})} & \frac{ -2i\tilde{J}}{4\tilde{J}^2+\omega^2 - i \omega k (v_{F_+}-v_{F_-})} & 0\\
0&\frac{ 2i\tilde{J}}{4\tilde{J}^2+\omega^2 - i \omega k (v_{F_+}-v_{F_-})}& \frac{i\omega + v_{F_+} k}{4\tilde{J}^2+\omega^2 - i \omega k (v_{F_+}-v_{F_-})} & 0 \\
\frac{ -2i\tilde{J}}{4\tilde{J}^2+\omega^2 + i \omega k (v_{F_+}-v_{F_-})}&0& 0& \frac{i\omega - v_{F_+} k}{4\tilde{J}^2+\omega^2 + i \omega k (v_{F_+}-v_{F_-})}
\end{pmatrix},
\end{equation}
\begin{equation}
	\mathbf{G}_m \approx \begin{pmatrix} 0 & 0 & 0 & \frac{i}{2\tilde{J}}\\
	0 & 0 & -\frac{i}{2\tilde{J}} & 0\\
	0&\frac{i}{2\tilde{J}}& 0 & 0 \\
	-\frac{i}{2\tilde{J}}&0& 0 & 0
	\end{pmatrix},
\end{equation} where we again expanded the Green's function using the smallness of \(\frac{t}{\tilde{J}}\). Using Eq.(\ref{eq:disgap}) gives for the effective action of the ungapped fermions

\begin{equation}
\mathcal{L}_{\rm dis} \simeq \frac{i g_{2k_{F_+}}(x)g_{2k_{F_-}}(x)}{4\tilde{J}}\left[\Rd_{+\downarrow}\Ll_{-\uparrow}+\Rd_{-\downarrow}\Ll_{+\uparrow}\right]+h.c.
\end{equation} Note, that the effective disorder mixes the band indices of the ungapped fermions. However, similar to the weak tunneling case, this effect is suppressed  \(g_{2k_{F_\pm}}/\tilde{J}\ll1\) and we expect ballistic transport of the gapless fermions up to parametrically large scales.

\section{Suppl.Mat. G: Dresselhaus Spin-Orbit interaction}

In the limit of vanishing tunneling, the KCs have almost equal ground state energies, in the sense that perturbative corrections to the uncoupled chains are small and we neglect them. Thus, there is a degenerate ground state. To find the spin configuration which wins in real materials, we introduce an additional Dresselhaus spin-orbit interaction, which is present for example in GaAs quantum wires.

\begin{equation}
\hat{H}_{\text{SOI}} = d \left(\ccd_{j} \hat{k} \operatorname{\sigma}_z \cc_{j}\right)^{(n)}, \quad n=1,2,
\end{equation} where \(\hat{k}\) is the momentum operator in the direction of the KCs. We assume, that the spin-orbit interaction is weak \(t_\perp \ll d < J \ll 2t\). We do not take into account the tunneling effect, since it is subleading. For the following calculations we set \(t_\perp=0\). The band structure of the non-interacting system in the presence of spin orbit interaction is given by

\begin{equation}
	\varepsilon_{\nicefrac{\uparrow}{\downarrow}} \! (k)= -2 t \cos \left(k  \xi_0 \right) \pm d  k,
\end{equation}

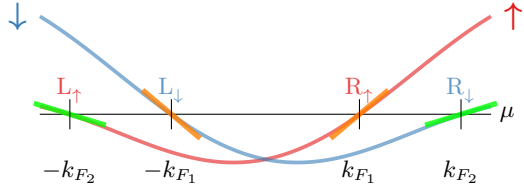
\begin{figure}[htbp]
	\input{figures/Soi}
	\caption{Bandstructure in the presence of Dresselhaus SOI \(\varepsilon_{\uparrow,\downarrow}=-2t \cos\left(k \xi_0\right) \pm d k\). The splitting due to the tunneling is subleading and not shown here. The slope of the linearized dispersion (orange and green), and thus the respective Fermi velocities, is different for \(\{\rm R_{\uparrow},L_{\downarrow}\}\) and \(\rm\{R_{\downarrow},L_{\uparrow}\}\). This allows us to distinguish the degenerate ground states in the weak tunneling limit.}
	\label{fig:soi}
\end{figure}

Note, that the spin-orbit interaction now effects different spins in different ways. This is the main difference to the splitting caused by the tunneling. We proceed similar to the case without SOI and place our chemical potential such that we find four Fermi points and single out smooth chiral modes in the following way

\begin{align}
	\cc_{\uparrow} &= e^{-ik_{F_1} x} \Rr_{\uparrow} + e^{i k_{F_2}x} \Ll_{\uparrow}, \\
	\cc_{\downarrow} &= e^{-ik_{F_2} x} \Rr_{\downarrow} + e^{i k_{F_1}x} \Ll_{\downarrow}.
\end{align} Note, that the Fermi velocities will be different for different helical sectors. \(\{\Rr_{\uparrow},\Ll_{\downarrow}\}\) depend on \(v_{F_1}= 2t \xi_0 \sin\left(k_{F_1} \xi_0 \right)+d\) and \(\{\Rr_{\downarrow},\Ll_{\uparrow}\}\) depend on \(v_{F_2}= 2t \xi_0 \sin\left(k_{F_2}\xi_0\right)-d\), respectively. The spin-orbit interaction is already diagonal in the wire space. We seperate the slow and fast spin degrees of freedom, following the steps presented in Suppl.Mat. B. We explicitly single out a fast \(\tilde{k}_{F_{12}}=\frac{1}{2}\left(k_{F_1}+k_{F_2}\right)\) component of the spins and assume, that the splitting \(\delta k_{F_{12}}=k_{F_1}-k_{F_2} = \frac{2}{\xi_0}\arcsin\left(\frac{d \tilde{k}_{F_{12}} \xi_0 }{2 \tilde{v}_{F_{12}}}\right)\) is small. The Kondo interaction is then given by

\begin{align}\label{eq:SOIgap1}
	\mathcal{L}^{h=+1}_{\rm bs } & = -\tilde{J}\left(\Rd \
	e^{-i\psi} \cos^2 \left(\frac{\theta}{2}\right) \sigma_+ \ \Ll\right)^{(n)} e^{i\delta k_{F_{12}}x} + h.c.\\\label{eq:SOIgap2}
	\mathcal{L}^{h=-1}_{\rm bs } & = \tilde{J}\left(\Rd \ e^{i\psi}\sin^2\left(\frac{\theta}{2}\right)  \sigma_- \ \Ll \right)^{(n)} e^{-i\delta k_{F_{12}}x}+h.c.
\end{align} which are two copies of the gap structure derived in \cite{oleg1,schimmel1}. The oscillations in Eqs. (\ref{eq:SOIgap1}) and (\ref{eq:SOIgap2}) are slow and can be gauged away. The gauge transformation leads to a small unimportant shift of the chemical, which we will not discuss here. In the following, we will ignore the oscillations. We can now compare the ground state energies of the two different spin configurations.\newline

{\noindent\bf Spin configuration  I: \(\theta^{(1)} = \theta^{(2)} = 0\)}\newline

The Green's function of the system is given by

\begin{equation}
- \mathbf{G}^{-1}=\begin{array}{c cccc |  cccc  c c   }
\ldelim({8}{1mm}&\partial_{R_1} & m & 0 & 0 & 0 &0 & 0 & 0 &\hspace{-.5cm} \rdelim){8}{-5mm} & R_{1\uparrow} \\
&m & \partial_{L_1} & 0 & 0 &0  & 0& 0 & 0 & & L_{1\downarrow} \\
&0 & 0 & \partial_{R_1} &m  & 0 & 0 & 0 &0 & &R_{2\uparrow} \\
&0 & 0 &m  & \partial_{L_1} & 0 & 0 &0 & 0 && L_{2\downarrow} \\ \cline{2-9}&&&&&&&&&\\[-10pt]
&0 &0 & 0 & 0 & \partial_{R_2} & 0  & 0 &  0 & &R_{1\downarrow} \\
&0 & 0 & 0 & 0 & 0 & \partial_{L_2} & 0 & 0 & &L_{1\uparrow} \\
&0 & 0 & 0 &0 & 0 & 0 & \partial_{R_2} & 0 & &R_{2\downarrow} \\
&0 & 0 &0 & 0 & 0 &0 & 0  & \partial_{L_2}  & &L_{2\uparrow}
\end{array},
\vspace{.5cm}
\end{equation}where \(m= 2 \tilde{J}\) and \(\partial_{\nicefrac{R}{L}l}=\partial_\tau \mp i v_{F_l} \partial_x, \ l=1,2 \). Note that the gap in the 1D case is twice as large as in the diagonalized basis. This is just a matter of the definition of \(\tilde{J}\). We use Eq.(\ref{eq:Gse}) from Supp. Mat. C and compute the ground state energy by integrating out the gapped fermions. We find

\begin{equation}\label{eq:Gsesoi}
	\delta E^{(\rm WT,I,SOI)} =  -\frac{2\xi_0 \tilde{J}^2}{\pi v_{F_1}}	\log\left(\frac{2t}{|\tilde{J}|}\right),
\end{equation} Since the spin-orbit interaction is weak, we can expand the Fermi velocity using the smallness of \(\frac{d}{t}\ll 1\). We find

\begin{equation}\label{eq:GsesoiI}
\delta E^{(\rm WT,I,SOI)} \approx  -\frac{2\xi_0 \tilde{J}^2}{\pi \left(\tilde{v}_{F_{12}}+d\right)}	\log\left(\frac{2t}{|\tilde{J}|}\right).
\end{equation} Note, that if we choose \(\theta^{(1)} = \theta^{(2)} = \pi\), Eq.(\ref{eq:Gsesoi}) would depend on \(\tilde{v}_{F_{12}}-d\) in the denominator. Let us assume, that \(d<0\). This means, that gapping all fermions with helicity \(h=+1\) is more favorable than gapping all fermions with helicity \(h=-1\). The resulting helical phase will thus always have gapless modes with helicity \(h=-1\).\newline

{\noindent \bf Spin configuration  II: \(\theta^{(1)} = 0, \, \theta^{(2)} = \pi\)}\newline

The Green's function of the system is given by

\begin{equation}
- \mathbf{G}^{-1}=\begin{array}{c cccc |  cccc  c c   }
\ldelim({8}{1mm}&\partial_{R_1} & m & 0 & 0 & 0 &0 & 0 & 0 &\hspace{-.5cm} \rdelim){8}{-5mm} & R_{1\uparrow} \\
&m & \partial_{L_1} & 0 & 0 &0  & 0& 0 & 0 & & L_{1\downarrow} \\
&0 & 0 & \partial_{R_1} &0  & 0 & 0 & 0 &0 & &R_{2\uparrow} \\
&0 & 0 &0  & \partial_{L_1} & 0 & 0 &0 & 0 && L_{2\downarrow} \\ \cline{2-9}&&&&&&&&&\\[-10pt]
&0 &0 & 0 & 0 & \partial_{R_2} & 0  & 0 &  0 & &R_{1\downarrow} \\
&0 & 0 & 0 & 0 & 0 & \partial_{L_2} & 0 & 0 & &L_{1\uparrow} \\
&0 & 0 & 0 &0 & 0 & 0 & \partial_{R_2} & m & &R_{2\downarrow} \\
&0 & 0 &0 & 0 & 0 &0 & m  & \partial_{L_2}  & &L_{2\uparrow}
\end{array},
\vspace{.5cm}
\end{equation} with \(m=2\tilde{J}\). We follow the same steps as before and find

\begin{equation}\label{eq:GsesoiII}
\delta E^{(\rm WT,II,SOI)} \approx  -\frac{\xi_0 \tilde{J}^2}{\pi}\left[\frac{1}{\tilde{v}_{F_{12}}+d}+\frac{1}{\tilde{v}_{F_{12}}-d}\right]	\log\left(\frac{2t}{|\tilde{J}|}\right).
\end{equation} If we now compare Eqs.(\ref{eq:GsesoiI}) and (\ref{eq:GsesoiII}) we can see, that for any choice of \(d\neq 0\) the fully helical phase will always be energetically more favorable. For \(d < 0\) the gapless modes have helicity \(h=-1\) and  for \(d > 0\) the gapless modes have helicity \(h=+1\).

\section{Suppl.Mat. H: Renormalization of the Luttinger parameter \(K_\psi\)}

\subsection{Intermediate tunneling}

After integrating out the gapped fermions we obtain the following low energy Lagrangian

\begin{equation}\label{eq:lowL1}
	\mathcal{L}=  \frac{\delta E}{\xi_0}+ \frac{1}{2\pi}\left[(\partial_\tau \tilde{\psi})^2 + (v_{F_+} \partial_x \tilde{\psi})^2\right] + \frac{1}{2\pi}\left[(\partial_\tau \tilde{\psi})^2 + (v_{F_-} \partial_x \tilde{\psi})^2\right]  +\mathcal{L}_{WZ}+\mathcal{L}_0[\Rr_{\pm\downarrow},\Ll_{\pm\uparrow}],
\end{equation} where \(\mathcal{L}_0\) is the chiral Lagrangian for the gapless fermions and  \(\mathcal{L}_{WZ}\) is the slow Wess Zumino term derived in \cite{oleg1,schimmel1}, which is given by

\begin{equation}
	\mathcal{L}_{WZ}= \frac{i s \rho_s}{\xi_0} \sin\left(\alpha^{(n)}\right)	\cos\left(\theta^{(n)}\right) \partial_\tau \psi^{(n)}.
\end{equation}

For spin configuration I the classical values are \(\alpha^{(n)}=0\) and \(\theta^{(n)}=0\) or \(\theta^{(n)}=\pi\). Let us assume the first case \(\theta^{(n)}=0\). The ground state energy equation in the intermediate tunneling regime is given by

\begin{equation}
	\frac{\delta E^{\rm (IT)}}{\xi_0} =  \underbrace{\frac{(s\rho_s \frac{J}{2})^2}{\pi \left(v_{F_+}+v_{F_-}\right)}\log\left(\frac{2t}{|\tilde{J}|}\right)}_{CJ^2}\left[-4+4\delta \psi'^2 + 4\left(\tilde{\alpha}^2+\delta\alpha^2\right) + 2\left(\tilde{\theta}^2+\delta\theta^2\right)  \right];
\end{equation}where \(\delta\psi'=\delta\psi-\frac{\pi}{2}\),  \(\nicefrac{\tilde{\alpha}}{\tilde{\theta}}=\frac{1}{2}\left(\nicefrac{\alpha}{\theta}^{(1)}+\nicefrac{\alpha}{\theta}^{(2)}\right)\) and \(\nicefrac{\delta\alpha}{\delta\theta}=\frac{1}{2}\left(\nicefrac{\alpha}{\theta}^{(1)}-\nicefrac{\alpha}{\theta}^{(2)}\right)\). In leading order the Wess Zumino term reads as

\begin{equation}\label{eq:WZ1}
\mathcal{L}_{WZ}= \underbrace{\frac{2 i s \rho_s}{\xi_0}}_{i D}\left[ \tilde{\alpha} \partial_\tau \tilde{\psi} +  \delta \alpha \partial_\tau \delta\psi'\right].
\end{equation}

Note, that the only massless field in our theory is \(\tilde{\psi}\). Integrating out the massive variables will thus lead to a renormalization of the compressibility and velocity of the Luttinger liquid action in (\ref{eq:lowL1}). Since we are interested only in the low frequency behavior of the system, we neglect the second term in (\ref{eq:WZ1}), because \(\delta\psi'\) is a massive variable \(\sim J^2\). After integrating out the massive fields, we obtain the following Lagrangian

\begin{equation}\label{eq:lowL2}
\mathcal{L}=  \frac{1}{2\pi}\left[(\partial_\tau \tilde{\psi})^2 + (v_{F_+} \partial_x \tilde{\psi})^2\right] + \frac{1}{2\pi}\left[(\partial_\tau \tilde{\psi})^2 + (v_{F_-} \partial_x \tilde{\psi})^2\right] + \frac{D^2}{16CJ^2}\left(\partial_\tau \tilde{\psi}\right)^2,
\end{equation} Slightly rewriting this Lagrangian gives the usual LL Lagrangian with renormalized compressibility \(K_\psi\)

\begin{equation}\label{eq:lowL3}
\mathcal{L}=  \frac{1}{K_\psi}\mathcal{L}[\tilde{\psi},v_\psi]+ \mathcal{L}_0[\Rr_{\pm\downarrow},\Ll_{\pm\uparrow}],
\end{equation} with \(\mathcal{L}[\tilde{\psi},v_\psi]=  \frac{1}{2\pi K_\psi}\left[(\partial_\tau \tilde{\psi})^2 + (v_{F_\psi} \partial_x \tilde{\psi})^2\right]\) and \(v_\psi= K_\psi\sqrt{v_{F_+}^2+v_{F_-}^2} \). The compressibility becomes strongly renormalized and is given by

\begin{equation}
	K_\psi=\sqrt{\frac{1}{2+\frac{D^2}{16C J^2}}} \approx \frac{4\sqrt{C}}{D}J \ll 1,
\end{equation} where we used the fact, that \(J/t\ll1\) and expanded in the last step. It was shown in \cite{oleg1,schimmel1}, that Eq.(\ref{eq:lowL3}), upon bosonization, consists of two helical U(1) Luttinger liquids, which couple to charge an spin sources simultaneously. The collective mode \(\tilde{\psi}\) becomes strongly renormalized.

\section{Suppl.Mat. I: Asymmetrically doped Kondo chains}

In the main text, we have considered the case of two identical Kondo chains with constant Kondo couplings, $ J_1 = J_2 = {\rm const} $. 
In more (experimentally) realistic situations, magnetic doping cannot be ideal and $ J_{1,2} $ fluctuate in space. Irregularity 
of the doping within one wire is not expected to destroy the helical phase with the protected transport \cite{oleg1,oleg2,oleg3,schimmel1}.
Let us show that the inter-chain fluctuations, $ J_1 \ne J_2 $, also do not change our conclusions, at least if the relative fluctuations are small.

We start with the Hamiltonian of two coupled KCs with the different Kondo couplings:
\begin{equation}
\hat{H}= \hat{H}_0+ \hat{H}_\perp + \sum_{\substack{f=1,2\\a=x,y,z}} J_{f,a}\operatorname{c}^\dagger_{f} S_f^a \sigma_a \operatorname{c}^{\vphantom{\dagger}}_{f}.
\end{equation}  
We have omitted summation over the position index \(j\) for better readability. We follow the calculations in the main text and change to the basis 
\(c_{1,2}=\frac{1}{\sqrt{2} }(\operatorname{c}_+ \pm \operatorname{c}_-)\) where \(\hat{H}_0+\hat{H}_\perp\) is diagonal. The interaction Hamiltonian is given by
\begin{multline}
\label{Ham_SM-I}
\hat{H}_{KI}=\frac{1}{2}\sum_a   (\operatorname{c}^\dagger_+ + \operatorname{c}^\dagger_-) J_{1,a} S_1^a \sigma_a (\operatorname{c}^{\vphantom{\dagger}}_+ + \operatorname{c}^{\vphantom{\dagger}}_-) + (\operatorname{c}^\dagger_+ - \operatorname{c}^\dagger_-) J_{2,a} S_2^a \sigma_a (\operatorname{c}^{\vphantom{\dagger}}_+ - \operatorname{c}^{\vphantom{\dagger}}_-)=\\
= \frac{1}{2}\sum_{a,\nu=\pm} \operatorname{c}^\dagger_{\nu} (J_{1,a} S_1^a + J_{2,a} S_2^a) \sigma_a\operatorname{c}^{\vphantom{\dagger}}_{\nu}+ \frac{1}{2}\sum_{a,\nu\neq\nu'} \operatorname{c}^\dagger_{\nu} (J_{1,a} S_1^a - J_{2,a} S_2^a) \sigma_a\operatorname{c}^{\vphantom{\dagger}}_{\nu'}.
\end{multline}
Using the mean value of the coupling constants and their fluctuation, \(\tilde{J}_{a}=\frac{1}{2}(J_{1,a}+J_{2,a})\) and \(\delta J_a = \frac{1}{2}(J_{1,a} - J_{2,a})\), and
composite spin variables, \(S^a_\pm = S_1^a \pm S_2^a\), we can rewrite Eq.(\ref{Ham_SM-I}):
%
%
\begin{equation}
\hat{H}_{KI}=\sum_{a,\nu} \operatorname{c}^\dagger_{\nu} \left(\tilde{J}_a S_+^a + \delta J_a  S_-^a\right) \sigma_a\operatorname{c}^{\vphantom{\dagger}}_{\nu}+ \sum_{a,\nu\neq\nu'} \operatorname{c}^\dagger_{\nu} \left(\tilde{J}_a S_-^a + \delta J_a  S_+^a\right) \sigma_a\operatorname{c}^{\vphantom{\dagger}}_{\nu'}.
\end{equation}
We continue in the Lagrangian formalism, linearize the spectrum of the electrons and introduce smooth left and right moving fermionic modes, $ \rm R, L $. 
This yields:
\begin{equation}
\label{Ham_SM-I-2}
\mathcal{L}_{KI}=\sum_{a,\nu} \operatorname{R}^\dagger_{\nu} \left(\tilde{J}_a S_+^a + \delta J_a  S_-^a\right) \sigma_a\operatorname{L}^{\vphantom{\dagger}}_{\nu} e^{2ik_F^\nu x} + \sum_{a,\nu\neq\nu'} \operatorname{R}^\dagger_{\nu} \left(\tilde{J}_a S_-^a + \delta J_a  S_+^a\right) \sigma_a\operatorname{L}^{\vphantom{\dagger}}_{\nu'} e^{i\left(k_F^++k_F^-\right)x} +h.c.
\end{equation}
Let us assume for simplicity that the $J-$fluctuations are small, \(\delta J/\tilde{J} \ll 1 \). Following the approach
described in the main text, one can now decompose the spins into fast and slow components and analyze the three regimes 
of strong, intermediate and weak tunneling. It is clear from Eq.(\ref{Ham_SM-I-2}) that the small fluctuations generate
inessential corrections to the fermionic gap whose relative smallness is controlled by the parameter $ \delta J/\tilde{J} $. 
Importantly, the number of the gaps is unaffected. We thus conclude that our prediction of the helical phase
in the coupled KCs remains unchanged.

%% file: figures/imspin.tex
\centering


\tdplotsetmaincoords{60}{110}

\pgfmathsetmacro{\rvec}{.8}
\pgfmathsetmacro{\thetavec}{30}
\pgfmathsetmacro{\phivec}{60}

\begin{tikzpicture}[scale=5,tdplot_main_coords]

\coordinate (O) at (0,0,0);

\tdplotsetcoord{P}{\rvec}{\thetavec}{\phivec}


\draw[thick,->] (0,0,0) -- (1,0,0) node[anchor=north east]{$1$};
\draw[thick,->] (0,0,0) -- (0,1,0) node[anchor=north west]{$2$};
\draw[thick,->] (0,0,0) -- (0,0,1) node[anchor=south]{$3$};
\draw[white,opacity=0] (0,0,0) -- (0,0,1.3);

\draw[-stealth,color=red,line width=.7mm] (O) -- (P) node[right,above,shift={(0,0.2)}] {$\mathbf{S}$};


\draw[color=red,line width=.7mm,-stealth] (O) -- (Pxy) node[near end, below] {$\mathbf{S}_{fast}$};
\draw[color=red,line width=.7mm,stealth-] ($ (P) - (Pxy) $) -- (O) node[near start, left] {$\mathbf{S}_{slow}$};
\draw[color=red, dashed] (P) -- ($ (O) + (P) - (Pxy) $);
\draw[color=red, dashed] (Pxy) -- (P);

\tdplotdrawarc[blue]{(O)}{0.15}{0}{\phivec}{anchor=north}{$\alpha$}

\tdplotsetthetaplanecoords{\phivec}

\tdplotdrawarc[tdplot_rotated_coords,blue]{(0,0,0)}{0.2}{90}{\thetavec}{anchor=south west}{$\alpha_\parallel$}

\draw[dashed,tdplot_rotated_coords] (\rvec,0,0) arc (0:90:\rvec);
\draw[dashed] (\rvec,0,0) arc (0:90:\rvec);

\tdplotsetcoord{Pp}{.5}{90}{\phivec}

\end{tikzpicture}

%% file: figures/Soi.tex
\centering

\begin{minipage}[c]{\linewidth}
	\centering
	\begin{tikzpicture}[scale=1]
	\draw (0.5,-0.2) -- (6.5,-0.2) node[right] {$\mu$};
	\begin{scope}[shift={(3.5,0)}]
		\draw[scale=1,domain=-3:3,smooth,variable=\x,fred
		,line width=0.5mm, samples=360,opacity=0.6] plot (\x,{-.75*cos(4/5*\x r)+0.2*\x-0.05}) node[right,opacity=1,scale=1.3] {$\boldsymbol{\uparrow}$};
		\draw[scale=1,domain=3:-3,smooth,variable=\x,fbrown
		,line width=0.5mm, samples=360,opacity=0.6] plot (\x,{-.75*cos(4/5*\x r)+-0.2*\x-0.05}) node[left,opacity=1,scale=1.3] {$\boldsymbol{ \downarrow}$};
		
		\begin{scope}[shift={(-1.25,-.2)}]
		\begin{scope}[rotate=50]
		\draw[line width=0.7mm,orange,opacity=0.8] (0,.5) -- (0,-.5) node[midway,above,fbrown] {$\Ll_{\downarrow}$};
		\end{scope}
		\end{scope}
		
		\begin{scope}[shift={(+1.25,-.2)}]
		\begin{scope}[rotate=-50]
		\draw[line width=0.7mm,orange,opacity=0.8] (0,.5) -- (0,-.5) node[midway,above,fred] {$\Rr_{\uparrow}$};
		\end{scope}
		\end{scope}
		
		\begin{scope}[shift={(+2.6,-.2)}]
		\begin{scope}[rotate=-73]
		\draw[line width=0.7mm,green,opacity=0.8] (0,.5) -- (0,-.5) node[midway,above,fbrown] {$\Rr_{\downarrow}$};
		\end{scope}
		\end{scope}
		
		\begin{scope}[shift={(-2.6,-.2)}]
		\begin{scope}[rotate=73]
		\draw[line width=0.7mm,green,opacity=0.8] (0,.5) -- (0,-.5) node[midway,above,fred] {$\Ll_{\uparrow}$};
		\end{scope}
		\end{scope}
		\draw (1.25,0) -- (1.25,-.4) node[below,shift={(0,-0.3)}] {$k_{F_1}$};
		\draw (-1.25,0) -- (-1.25,-.4) node[below,shift={(0,-0.3)}] {$-k_{F_1}$};
		\draw (2.6,0) -- (2.6,-.4) node[below,shift={(0,-0.3)}] {$k_{F_2}$};
		\draw (-2.6,0) -- (-2.6,-.4) node[below,shift={(0,-0.3)}] {$-k_{F_2}$};

	\end{scope}
	
	\end{tikzpicture}
\end{minipage} 

%% file: main.bbl
\begin{thebibliography}{57}%
	\makeatletter
	\providecommand \@ifxundefined [1]{%
		\@ifx{#1\undefined}
	}%
	\providecommand \@ifnum [1]{%
		\ifnum #1\expandafter \@firstoftwo
		\else \expandafter \@secondoftwo
		\fi
	}%
	\providecommand \@ifx [1]{%
		\ifx #1\expandafter \@firstoftwo
		\else \expandafter \@secondoftwo
		\fi
	}%
	\providecommand \natexlab [1]{#1}%
	\providecommand \enquote  [1]{``#1''}%
	\providecommand \bibnamefont  [1]{#1}%
	\providecommand \bibfnamefont [1]{#1}%
	\providecommand \citenamefont [1]{#1}%
	\providecommand \href@noop [0]{\@secondoftwo}%
	\providecommand \href [0]{\begingroup \@sanitize@url \@href}%
	\providecommand \@href[1]{\@@startlink{#1}\@@href}%
	\providecommand \@@href[1]{\endgroup#1\@@endlink}%
	\providecommand \@sanitize@url [0]{\catcode `\\12\catcode `\$12\catcode
		`\&12\catcode `\#12\catcode `\^12\catcode `\_12\catcode `\%12\relax}%
	\providecommand \@@startlink[1]{}%
	\providecommand \@@endlink[0]{}%
	\providecommand \url  [0]{\begingroup\@sanitize@url \@url }%
	\providecommand \@url [1]{\endgroup\@href {#1}{\urlprefix }}%
	\providecommand \urlprefix  [0]{URL }%
	\providecommand \Eprint [0]{\href }%
	\providecommand \doibase [0]{http://dx.doi.org/}%
	\providecommand \selectlanguage [0]{\@gobble}%
	\providecommand \bibinfo  [0]{\@secondoftwo}%
	\providecommand \bibfield  [0]{\@secondoftwo}%
	\providecommand \translation [1]{[#1]}%
	\providecommand \BibitemOpen [0]{}%
	\providecommand \bibitemStop [0]{}%
	\providecommand \bibitemNoStop [0]{.\EOS\space}%
	\providecommand \EOS [0]{\spacefactor3000\relax}%
	\providecommand \BibitemShut  [1]{\csname bibitem#1\endcsname}%
	\let\auto@bib@innerbib\@empty
	\bibitem [{\citenamefont {Giamarchi}(2003)}]{Giamarchi}%
	\BibitemOpen
	\bibfield  {author} {\bibinfo {author} {\bibfnamefont {T.}~\bibnamefont
			{Giamarchi}},\ }\href
	{https://www.oxfordscholarship.com/view/10.1093/acprof:oso/9780198525004.001.0001/acprof-9780198525004}
	{\emph {\bibinfo {title} {Quantum {Physics} in {One} {Dimension}}}}\
	(\bibinfo  {publisher} {Oxford University Press},\ \bibinfo {year}
	{2003})\BibitemShut {NoStop}%
	\bibitem [{\citenamefont {Shen}(2017)}]{TI3}%
	\BibitemOpen
	\bibfield  {author} {\bibinfo {author} {\bibfnamefont {S.-Q.}\ \bibnamefont
			{Shen}},\ }\href {\doibase 10.1007/978-981-10-4606-3} {\emph {\bibinfo
			{title} {Topological {Insulators}: {Dirac} {Equation} in {Condensed}
				{Matter}}}}\ (\bibinfo  {publisher} {Springer Singapore},\ \bibinfo {year}
	{2017})\BibitemShut {NoStop}%
	\bibitem [{\citenamefont {Hasan}\ and\ \citenamefont {Kane}(2010)}]{TI1}%
	\BibitemOpen
	\bibfield  {author} {\bibinfo {author} {\bibfnamefont {M.~Z.}\ \bibnamefont
			{Hasan}}\ and\ \bibinfo {author} {\bibfnamefont {C.~L.}\ \bibnamefont
			{Kane}},\ }\href {\doibase 10.1103/RevModPhys.82.3045} {\bibfield  {journal}
		{\bibinfo  {journal} {Rev. Mod. Phys.}\ }\textbf {\bibinfo {volume} {82}},\
		\bibinfo {pages} {3045–3067} (\bibinfo {year} {2010})}\BibitemShut
	{NoStop}%
	\bibitem [{\citenamefont {Qi}\ and\ \citenamefont {Zhang}(2011)}]{topin}%
	\BibitemOpen
	\bibfield  {author} {\bibinfo {author} {\bibfnamefont {X.-L.}\ \bibnamefont
			{Qi}}\ and\ \bibinfo {author} {\bibfnamefont {S.-C.}\ \bibnamefont {Zhang}},\
	}\href {\doibase 10.1103/RevModPhys.83.1057} {\bibfield  {journal} {\bibinfo
			{journal} {Rev. Mod. Phys.}\ }\textbf {\bibinfo {volume} {83}},\ \bibinfo
		{pages} {1057–1110} (\bibinfo {year} {2011})}\BibitemShut {NoStop}%
	\bibitem [{\citenamefont {Bernevig}\ and\ \citenamefont
		{Zhang}(2006)}]{BenZhang}%
	\BibitemOpen
	\bibfield  {author} {\bibinfo {author} {\bibfnamefont {B.~A.}\ \bibnamefont
			{Bernevig}}\ and\ \bibinfo {author} {\bibfnamefont {S.-C.}\ \bibnamefont
			{Zhang}},\ }\href {\doibase 10.1103/PhysRevLett.96.106802} {\bibfield
		{journal} {\bibinfo  {journal} {Phys. Rev. Lett.}\ }\textbf {\bibinfo
			{volume} {96}},\ \bibinfo {pages} {106802} (\bibinfo {year}
		{2006})}\BibitemShut {NoStop}%
	\bibitem [{\citenamefont {Bernevig}\ \emph {et~al.}(2006)\citenamefont
		{Bernevig}, \citenamefont {Hughes},\ and\ \citenamefont {Zhang}}]{BenZhang2}%
	\BibitemOpen
	\bibfield  {author} {\bibinfo {author} {\bibfnamefont {B.~A.}\ \bibnamefont
			{Bernevig}}, \bibinfo {author} {\bibfnamefont {T.~L.}\ \bibnamefont
			{Hughes}}, \ and\ \bibinfo {author} {\bibfnamefont {S.-C.}\ \bibnamefont
			{Zhang}},\ }\href {\doibase 10.1126/science.1133734} {\bibfield  {journal}
		{\bibinfo  {journal} {Science}\ }\textbf {\bibinfo {volume} {314}},\ \bibinfo
		{pages} {1757} (\bibinfo {year} {2006})}\BibitemShut {NoStop}%
	\bibitem [{\citenamefont {Liu}\ \emph {et~al.}(2008)\citenamefont {Liu},
		\citenamefont {Hughes}, \citenamefont {Qi}, \citenamefont {Wang},\ and\
		\citenamefont {Zhang}}]{AssymQ-Wells}%
	\BibitemOpen
	\bibfield  {author} {\bibinfo {author} {\bibfnamefont {C.}~\bibnamefont
			{Liu}}, \bibinfo {author} {\bibfnamefont {T.~L.}\ \bibnamefont {Hughes}},
		\bibinfo {author} {\bibfnamefont {X.-L.}\ \bibnamefont {Qi}}, \bibinfo
		{author} {\bibfnamefont {K.}~\bibnamefont {Wang}}, \ and\ \bibinfo {author}
		{\bibfnamefont {S.-C.}\ \bibnamefont {Zhang}},\ }\href {\doibase
		10.1103/PhysRevLett.100.236601} {\bibfield  {journal} {\bibinfo  {journal}
			{Phys. Rev. Lett.}\ }\textbf {\bibinfo {volume} {100}},\ \bibinfo {pages}
		{236601} (\bibinfo {year} {2008})}\BibitemShut {NoStop}%
	\bibitem [{\citenamefont {Drozdov}\ \emph {et~al.}(2014)\citenamefont
		{Drozdov}, \citenamefont {Alexandradinata}, \citenamefont {Jeon},
		\citenamefont {Nadj-Perge}, \citenamefont {Ji}, \citenamefont {Cava},
		\citenamefont {Bernevig},\ and\ \citenamefont {Yazdani}}]{BismuthTI-Meas}%
	\BibitemOpen
	\bibfield  {author} {\bibinfo {author} {\bibfnamefont {I.~K.}\ \bibnamefont
			{Drozdov}}, \bibinfo {author} {\bibfnamefont {A.}~\bibnamefont
			{Alexandradinata}}, \bibinfo {author} {\bibfnamefont {S.}~\bibnamefont
			{Jeon}}, \bibinfo {author} {\bibfnamefont {S.}~\bibnamefont {Nadj-Perge}},
		\bibinfo {author} {\bibfnamefont {H.}~\bibnamefont {Ji}}, \bibinfo {author}
		{\bibfnamefont {R.~J.}\ \bibnamefont {Cava}}, \bibinfo {author}
		{\bibfnamefont {B.~A.}\ \bibnamefont {Bernevig}}, \ and\ \bibinfo {author}
		{\bibfnamefont {A.}~\bibnamefont {Yazdani}},\ }\href {\doibase
		10.1038/nphys3048} {\bibfield  {journal} {\bibinfo  {journal} {Nature
				Physics}\ }\textbf {\bibinfo {volume} {10}},\ \bibinfo {pages} {664–669}
		(\bibinfo {year} {2014})}\BibitemShut {NoStop}%
	\bibitem [{\citenamefont {Murani}\ \emph {et~al.}(2017)\citenamefont {Murani},
		\citenamefont {Kasumov}, \citenamefont {Sengupta}, \citenamefont {Kasumov},
		\citenamefont {Volkov}, \citenamefont {Khodos}, \citenamefont {Brisset},
		\citenamefont {Delagrange}, \citenamefont {Chepelianskii}, \citenamefont
		{Deblock}, \citenamefont {Bouchiat},\ and\ \citenamefont
		{Guéron}}]{murani_ballistic_2017BI}%
	\BibitemOpen
	\bibfield  {author} {\bibinfo {author} {\bibfnamefont {A.}~\bibnamefont
			{Murani}}, \bibinfo {author} {\bibfnamefont {A.}~\bibnamefont {Kasumov}},
		\bibinfo {author} {\bibfnamefont {S.}~\bibnamefont {Sengupta}}, \bibinfo
		{author} {\bibfnamefont {Y.~A.}\ \bibnamefont {Kasumov}}, \bibinfo {author}
		{\bibfnamefont {V.~T.}\ \bibnamefont {Volkov}}, \bibinfo {author}
		{\bibfnamefont {I.~I.}\ \bibnamefont {Khodos}}, \bibinfo {author}
		{\bibfnamefont {F.}~\bibnamefont {Brisset}}, \bibinfo {author} {\bibfnamefont
			{R.}~\bibnamefont {Delagrange}}, \bibinfo {author} {\bibfnamefont
			{A.}~\bibnamefont {Chepelianskii}}, \bibinfo {author} {\bibfnamefont
			{R.}~\bibnamefont {Deblock}}, \bibinfo {author} {\bibfnamefont
			{H.}~\bibnamefont {Bouchiat}}, \ and\ \bibinfo {author} {\bibfnamefont
			{S.}~\bibnamefont {Guéron}},\ }\href {\doibase 10.1038/ncomms15941}
	{\bibfield  {journal} {\bibinfo  {journal} {Nature Communications}\ }\textbf
		{\bibinfo {volume} {8}},\ \bibinfo {pages} {1} (\bibinfo {year}
		{2017})}\BibitemShut {NoStop}%
	\bibitem [{\citenamefont {Schindler}\ \emph {et~al.}(2018)\citenamefont
		{Schindler}, \citenamefont {Wang}, \citenamefont {Vergniory}, \citenamefont
		{Cook}, \citenamefont {Murani}, \citenamefont {Sengupta}, \citenamefont
		{Kasumov}, \citenamefont {Deblock}, \citenamefont {Jeon}, \citenamefont
		{Drozdov}, \citenamefont {Bouchiat}, \citenamefont {Guéron}, \citenamefont
		{Yazdani}, \citenamefont {Bernevig},\ and\ \citenamefont
		{Neupert}}]{schindler_higher-order_2018}%
	\BibitemOpen
	\bibfield  {author} {\bibinfo {author} {\bibfnamefont {F.}~\bibnamefont
			{Schindler}}, \bibinfo {author} {\bibfnamefont {Z.}~\bibnamefont {Wang}},
		\bibinfo {author} {\bibfnamefont {M.~G.}\ \bibnamefont {Vergniory}}, \bibinfo
		{author} {\bibfnamefont {A.~M.}\ \bibnamefont {Cook}}, \bibinfo {author}
		{\bibfnamefont {A.}~\bibnamefont {Murani}}, \bibinfo {author} {\bibfnamefont
			{S.}~\bibnamefont {Sengupta}}, \bibinfo {author} {\bibfnamefont {A.~Y.}\
			\bibnamefont {Kasumov}}, \bibinfo {author} {\bibfnamefont {R.}~\bibnamefont
			{Deblock}}, \bibinfo {author} {\bibfnamefont {S.}~\bibnamefont {Jeon}},
		\bibinfo {author} {\bibfnamefont {I.}~\bibnamefont {Drozdov}}, \bibinfo
		{author} {\bibfnamefont {H.}~\bibnamefont {Bouchiat}}, \bibinfo {author}
		{\bibfnamefont {S.}~\bibnamefont {Guéron}}, \bibinfo {author} {\bibfnamefont
			{A.}~\bibnamefont {Yazdani}}, \bibinfo {author} {\bibfnamefont {B.~A.}\
			\bibnamefont {Bernevig}}, \ and\ \bibinfo {author} {\bibfnamefont
			{T.}~\bibnamefont {Neupert}},\ }\href {\doibase 10.1038/s41567-018-0224-7}
	{\bibfield  {journal} {\bibinfo  {journal} {Nature Physics}\ }\textbf
		{\bibinfo {volume} {14}},\ \bibinfo {pages} {918} (\bibinfo {year}
		{2018})}\BibitemShut {NoStop}%
	\bibitem [{\citenamefont {Knez}\ \emph {et~al.}(2011)\citenamefont {Knez},
		\citenamefont {Du},\ and\ \citenamefont {Sullivan}}]{EdgeTransport-Exp0}%
	\BibitemOpen
	\bibfield  {author} {\bibinfo {author} {\bibfnamefont {I.}~\bibnamefont
			{Knez}}, \bibinfo {author} {\bibfnamefont {R.-R.}\ \bibnamefont {Du}}, \ and\
		\bibinfo {author} {\bibfnamefont {G.}~\bibnamefont {Sullivan}},\ }\href
	{\doibase 10.1103/PhysRevLett.107.136603} {\bibfield  {journal} {\bibinfo
			{journal} {\prl}\ }\textbf {\bibinfo {volume} {107}},\ \bibinfo {pages}
		{136603} (\bibinfo {year} {2011})}\BibitemShut {NoStop}%
	\bibitem [{\citenamefont {Knez}\ \emph {et~al.}(2014)\citenamefont {Knez},
		\citenamefont {Rettner}, \citenamefont {Yang}, \citenamefont {Parkin},
		\citenamefont {Du}, \citenamefont {Du},\ and\ \citenamefont
		{Sullivan}}]{EdgeTransport-Exp1}%
	\BibitemOpen
	\bibfield  {author} {\bibinfo {author} {\bibfnamefont {I.}~\bibnamefont
			{Knez}}, \bibinfo {author} {\bibfnamefont {C.~T.}\ \bibnamefont {Rettner}},
		\bibinfo {author} {\bibfnamefont {S.-H.}\ \bibnamefont {Yang}}, \bibinfo
		{author} {\bibfnamefont {S.~S.~P.}\ \bibnamefont {Parkin}}, \bibinfo {author}
		{\bibfnamefont {L.~J.}\ \bibnamefont {Du}}, \bibinfo {author} {\bibfnamefont
			{R.~R.}\ \bibnamefont {Du}}, \ and\ \bibinfo {author} {\bibfnamefont
			{G.}~\bibnamefont {Sullivan}},\ }\href {\doibase
		10.1103/PhysRevLett.112.026602} {\bibfield  {journal} {\bibinfo  {journal}
			{\prl}\ }\textbf {\bibinfo {volume} {112}},\ \bibinfo {pages} {026602}
		(\bibinfo {year} {2014})}\BibitemShut {NoStop}%
	\bibitem [{\citenamefont {Spanton}\ \emph {et~al.}(2014)\citenamefont
		{Spanton}, \citenamefont {Nowack}, \citenamefont {Du}, \citenamefont
		{Sullivan}, \citenamefont {Du},\ and\ \citenamefont
		{Moler}}]{EdgeTransport-Exp2}%
	\BibitemOpen
	\bibfield  {author} {\bibinfo {author} {\bibfnamefont {E.~M.}\ \bibnamefont
			{Spanton}}, \bibinfo {author} {\bibfnamefont {K.~C.}\ \bibnamefont {Nowack}},
		\bibinfo {author} {\bibfnamefont {L.~J.}\ \bibnamefont {Du}}, \bibinfo
		{author} {\bibfnamefont {G.}~\bibnamefont {Sullivan}}, \bibinfo {author}
		{\bibfnamefont {R.~R.}\ \bibnamefont {Du}}, \ and\ \bibinfo {author}
		{\bibfnamefont {K.~A.}\ \bibnamefont {Moler}},\ }\href {\doibase
		10.1103/PhysRevLett.113.026804} {\bibfield  {journal} {\bibinfo  {journal}
			{\prl}\ }\textbf {\bibinfo {volume} {113}},\ \bibinfo {pages} {026804}
		(\bibinfo {year} {2014})}\BibitemShut {NoStop}%
	\bibitem [{\citenamefont {K{\"o}nig}\ \emph {et~al.}(2007)\citenamefont
		{K{\"o}nig}, \citenamefont {Wiedmann}, \citenamefont {Br{\"u}ne},
		\citenamefont {Roth}, \citenamefont {Buhmann}, \citenamefont {Molenkamp},
		\citenamefont {Qi},\ and\ \citenamefont {Zhang}}]{Koenig}%
	\BibitemOpen
	\bibfield  {author} {\bibinfo {author} {\bibfnamefont {M.}~\bibnamefont
			{K{\"o}nig}}, \bibinfo {author} {\bibfnamefont {S.}~\bibnamefont {Wiedmann}},
		\bibinfo {author} {\bibfnamefont {C.}~\bibnamefont {Br{\"u}ne}}, \bibinfo
		{author} {\bibfnamefont {A.}~\bibnamefont {Roth}}, \bibinfo {author}
		{\bibfnamefont {H.}~\bibnamefont {Buhmann}}, \bibinfo {author} {\bibfnamefont
			{L.~W.}\ \bibnamefont {Molenkamp}}, \bibinfo {author} {\bibfnamefont {X.-L.}\
			\bibnamefont {Qi}}, \ and\ \bibinfo {author} {\bibfnamefont {S.-C.}\
			\bibnamefont {Zhang}},\ }\href {\doibase 10.1126/science.1148047} {\bibfield
		{journal} {\bibinfo  {journal} {Science}\ }\textbf {\bibinfo {volume}
			{318}},\ \bibinfo {pages} {766} (\bibinfo {year} {2007})}\BibitemShut
	{NoStop}%
	\bibitem [{\citenamefont {K{\"o}nig}\ \emph {et~al.}(2008)\citenamefont
		{K{\"o}nig}, \citenamefont {Buhmann}, \citenamefont {{W. Molenkamp}},
		\citenamefont {Hughes}, \citenamefont {Liu}, \citenamefont {Qi},\ and\
		\citenamefont {Zhang}}]{Koenig2}%
	\BibitemOpen
	\bibfield  {author} {\bibinfo {author} {\bibfnamefont {M.}~\bibnamefont
			{K{\"o}nig}}, \bibinfo {author} {\bibfnamefont {H.}~\bibnamefont {Buhmann}},
		\bibinfo {author} {\bibfnamefont {L.}~\bibnamefont {{W. Molenkamp}}},
		\bibinfo {author} {\bibfnamefont {T.}~\bibnamefont {Hughes}}, \bibinfo
		{author} {\bibfnamefont {C.-X.}\ \bibnamefont {Liu}}, \bibinfo {author}
		{\bibfnamefont {X.-L.}\ \bibnamefont {Qi}}, \ and\ \bibinfo {author}
		{\bibfnamefont {S.-C.}\ \bibnamefont {Zhang}},\ }\href {\doibase
		10.1143/JPSJ.77.031007} {\bibfield  {journal} {\bibinfo  {journal} {Journal
				of the Physical Society of Japan}\ }\textbf {\bibinfo {volume} {77}},\
		\bibinfo {pages} {031007} (\bibinfo {year} {2008})}\BibitemShut {NoStop}%
	\bibitem [{\citenamefont {V{\"a}yrynen}\ \emph {et~al.}(2016)\citenamefont
		{V{\"a}yrynen}, \citenamefont {Geissler},\ and\ \citenamefont
		{Glazman}}]{vayrynen_2016}%
	\BibitemOpen
	\bibfield  {author} {\bibinfo {author} {\bibfnamefont {J.~I.}\ \bibnamefont
			{V{\"a}yrynen}}, \bibinfo {author} {\bibfnamefont {F.}~\bibnamefont
			{Geissler}}, \ and\ \bibinfo {author} {\bibfnamefont {L.~I.}\ \bibnamefont
			{Glazman}},\ }\href {\doibase 10.1103/PhysRevB.93.241301} {\bibfield
		{journal} {\bibinfo  {journal} {Phys. Rev. B}\ }\textbf {\bibinfo {volume}
			{93}},\ \bibinfo {pages} {241301} (\bibinfo {year} {2016})}\BibitemShut
	{NoStop}%
	\bibitem [{\citenamefont {Altshuler}\ \emph {et~al.}(2013)\citenamefont
		{Altshuler}, \citenamefont {Aleiner},\ and\ \citenamefont {Yudson}}]{yud}%
	\BibitemOpen
	\bibfield  {author} {\bibinfo {author} {\bibfnamefont {B.~L.}\ \bibnamefont
			{Altshuler}}, \bibinfo {author} {\bibfnamefont {I.~L.}\ \bibnamefont
			{Aleiner}}, \ and\ \bibinfo {author} {\bibfnamefont {V.~I.}\ \bibnamefont
			{Yudson}},\ }\href {\doibase 10.1103/PhysRevLett.111.086401} {\bibfield
		{journal} {\bibinfo  {journal} {Phys. Rev. Lett.}\ }\textbf {\bibinfo
			{volume} {111}},\ \bibinfo {pages} {086401} (\bibinfo {year}
		{2013})}\BibitemShut {NoStop}%
	\bibitem [{\citenamefont {Yevtushenko}\ \emph {et~al.}(2015)\citenamefont
		{Yevtushenko}, \citenamefont {Wugalter}, \citenamefont {Yudson},\ and\
		\citenamefont {Altshuler}}]{oleg4}%
	\BibitemOpen
	\bibfield  {author} {\bibinfo {author} {\bibfnamefont {O.~M.}\ \bibnamefont
			{Yevtushenko}}, \bibinfo {author} {\bibfnamefont {A.}~\bibnamefont
			{Wugalter}}, \bibinfo {author} {\bibfnamefont {V.~I.}\ \bibnamefont
			{Yudson}}, \ and\ \bibinfo {author} {\bibfnamefont {B.~L.}\ \bibnamefont
			{Altshuler}},\ }\href
	{https://iopscience.iop.org/article/10.1209/0295-5075/112/57003} {\bibfield
		{journal} {\bibinfo  {journal} {EPL}\ }\textbf {\bibinfo {volume} {112}}
		(\bibinfo {year} {2015})}\BibitemShut {NoStop}%
	\bibitem [{\citenamefont {Yevtushenko}\ and\ \citenamefont
		{Yudson}(2019)}]{olegyud}%
	\BibitemOpen
	\bibfield  {author} {\bibinfo {author} {\bibfnamefont {O.~M.}\ \bibnamefont
			{Yevtushenko}}\ and\ \bibinfo {author} {\bibfnamefont {V.~I.}\ \bibnamefont
			{Yudson}},\ }\href {http://arxiv.org/abs/1909.08460} {\bibfield  {journal}
		{\bibinfo  {journal} {arXiv:1909.08460}\ } (\bibinfo {year}
		{2019})}\BibitemShut {NoStop}%
	\bibitem [{\citenamefont {Braunecker}\ \emph
		{et~al.}(2009{\natexlab{a}})\citenamefont {Braunecker}, \citenamefont
		{Simon},\ and\ \citenamefont {Loss}}]{Loss}%
	\BibitemOpen
	\bibfield  {author} {\bibinfo {author} {\bibfnamefont {B.}~\bibnamefont
			{Braunecker}}, \bibinfo {author} {\bibfnamefont {P.}~\bibnamefont {Simon}}, \
		and\ \bibinfo {author} {\bibfnamefont {D.}~\bibnamefont {Loss}},\ }\href
	{\doibase 10.1103/PhysRevLett.102.116403} {\bibfield  {journal} {\bibinfo
			{journal} {Phys. Rev. Lett.}\ }\textbf {\bibinfo {volume} {102}},\ \bibinfo
		{pages} {116403} (\bibinfo {year} {2009}{\natexlab{a}})}\BibitemShut
	{NoStop}%
	\bibitem [{\citenamefont {Klinovaja}\ \emph {et~al.}(2013)\citenamefont
		{Klinovaja}, \citenamefont {Stano}, \citenamefont {Yazdani},\ and\
		\citenamefont {Loss}}]{klin}%
	\BibitemOpen
	\bibfield  {author} {\bibinfo {author} {\bibfnamefont {J.}~\bibnamefont
			{Klinovaja}}, \bibinfo {author} {\bibfnamefont {P.}~\bibnamefont {Stano}},
		\bibinfo {author} {\bibfnamefont {A.}~\bibnamefont {Yazdani}}, \ and\
		\bibinfo {author} {\bibfnamefont {D.}~\bibnamefont {Loss}},\ }\href {\doibase
		10.1103/PhysRevLett.111.186805} {\bibfield  {journal} {\bibinfo  {journal}
			{Phys. Rev. Lett.}\ }\textbf {\bibinfo {volume} {111}},\ \bibinfo {pages}
		{186805} (\bibinfo {year} {2013})}\BibitemShut {NoStop}%
	\bibitem [{\citenamefont {Braunecker}\ \emph
		{et~al.}(2009{\natexlab{b}})\citenamefont {Braunecker}, \citenamefont
		{Simon},\ and\ \citenamefont {Loss}}]{loss2}%
	\BibitemOpen
	\bibfield  {author} {\bibinfo {author} {\bibfnamefont {B.}~\bibnamefont
			{Braunecker}}, \bibinfo {author} {\bibfnamefont {P.}~\bibnamefont {Simon}}, \
		and\ \bibinfo {author} {\bibfnamefont {D.}~\bibnamefont {Loss}},\ }\href
	{\doibase 10.1103/PhysRevB.80.165119} {\bibfield  {journal} {\bibinfo
			{journal} {Phys. Rev. B}\ }\textbf {\bibinfo {volume} {80}},\ \bibinfo
		{pages} {165119} (\bibinfo {year} {2009}{\natexlab{b}})}\BibitemShut
	{NoStop}%
	\bibitem [{\citenamefont {Klinovaja}\ \emph {et~al.}(2012)\citenamefont
		{Klinovaja}, \citenamefont {Ferreira},\ and\ \citenamefont
		{Loss}}]{CNTklinloss}%
	\BibitemOpen
	\bibfield  {author} {\bibinfo {author} {\bibfnamefont {J.}~\bibnamefont
			{Klinovaja}}, \bibinfo {author} {\bibfnamefont {G.~J.}\ \bibnamefont
			{Ferreira}}, \ and\ \bibinfo {author} {\bibfnamefont {D.}~\bibnamefont
			{Loss}},\ }\href {\doibase 10.1103/PhysRevB.86.235416} {\bibfield  {journal}
		{\bibinfo  {journal} {Phys. Rev. B}\ }\textbf {\bibinfo {volume} {86}},\
		\bibinfo {pages} {235416} (\bibinfo {year} {2012})}\BibitemShut {NoStop}%
	\bibitem [{\citenamefont {Klinovaja}\ \emph {et~al.}(2011)\citenamefont
		{Klinovaja}, \citenamefont {Schmidt}, \citenamefont {Braunecker},\ and\
		\citenamefont {Loss}}]{CNTklinloss2}%
	\BibitemOpen
	\bibfield  {author} {\bibinfo {author} {\bibfnamefont {J.}~\bibnamefont
			{Klinovaja}}, \bibinfo {author} {\bibfnamefont {M.~J.}\ \bibnamefont
			{Schmidt}}, \bibinfo {author} {\bibfnamefont {B.}~\bibnamefont {Braunecker}},
		\ and\ \bibinfo {author} {\bibfnamefont {D.}~\bibnamefont {Loss}},\ }\href
	{\doibase 10.1103/PhysRevLett.106.156809} {\bibfield  {journal} {\bibinfo
			{journal} {Phys. Rev. Lett.}\ }\textbf {\bibinfo {volume} {106}},\ \bibinfo
		{pages} {156809} (\bibinfo {year} {2011})}\BibitemShut {NoStop}%
	\bibitem [{\citenamefont {Quay}\ \emph {et~al.}(2010)\citenamefont {Quay},
		\citenamefont {Hughes}, \citenamefont {Sulpizio}, \citenamefont {Pfeiffer},
		\citenamefont {Baldwin}, \citenamefont {West}, \citenamefont
		{Goldhaber-Gordon},\ and\ \citenamefont {de~Picciotto}}]{quay}%
	\BibitemOpen
	\bibfield  {author} {\bibinfo {author} {\bibfnamefont {C.~H.~L.}\
			\bibnamefont {Quay}}, \bibinfo {author} {\bibfnamefont {T.~L.}\ \bibnamefont
			{Hughes}}, \bibinfo {author} {\bibfnamefont {J.~A.}\ \bibnamefont
			{Sulpizio}}, \bibinfo {author} {\bibfnamefont {L.~N.}\ \bibnamefont
			{Pfeiffer}}, \bibinfo {author} {\bibfnamefont {K.~W.}\ \bibnamefont
			{Baldwin}}, \bibinfo {author} {\bibfnamefont {K.~W.}\ \bibnamefont {West}},
		\bibinfo {author} {\bibfnamefont {D.}~\bibnamefont {Goldhaber-Gordon}}, \
		and\ \bibinfo {author} {\bibfnamefont {R.}~\bibnamefont {de~Picciotto}},\
	}\href {\doibase 10.1038/nphys1626} {\bibfield  {journal} {\bibinfo
			{journal} {Nature Physics}\ }\textbf {\bibinfo {volume} {6}},\ \bibinfo
		{pages} {336} (\bibinfo {year} {2010})}\BibitemShut {NoStop}%
	\bibitem [{\citenamefont {Heedt}\ \emph {et~al.}(2017)\citenamefont {Heedt},
		\citenamefont {{Traverso Ziani}}, \citenamefont {Cr{\'e}pin}, \citenamefont
		{Prost}, \citenamefont {Trellenkamp}, \citenamefont {Schubert}, \citenamefont
		{Gr{\"u}tzmacher}, \citenamefont {Trauzettel},\ and\ \citenamefont
		{Sch{\"a}pers}}]{heedt}%
	\BibitemOpen
	\bibfield  {author} {\bibinfo {author} {\bibfnamefont {S.}~\bibnamefont
			{Heedt}}, \bibinfo {author} {\bibfnamefont {N.}~\bibnamefont {{Traverso
					Ziani}}}, \bibinfo {author} {\bibfnamefont {F.}~\bibnamefont {Cr{\'e}pin}},
		\bibinfo {author} {\bibfnamefont {W.}~\bibnamefont {Prost}}, \bibinfo
		{author} {\bibfnamefont {S.}~\bibnamefont {Trellenkamp}}, \bibinfo {author}
		{\bibfnamefont {J.}~\bibnamefont {Schubert}}, \bibinfo {author}
		{\bibfnamefont {D.}~\bibnamefont {Gr{\"u}tzmacher}}, \bibinfo {author}
		{\bibfnamefont {B.}~\bibnamefont {Trauzettel}}, \ and\ \bibinfo {author}
		{\bibfnamefont {T.}~\bibnamefont {Sch{\"a}pers}},\ }\href
	{https://doi.org/10.1038/nphys4070} {\bibfield  {journal} {\bibinfo
			{journal} {Nature Physics}\ }\textbf {\bibinfo {volume} {13}},\ \bibinfo
		{pages} {563} (\bibinfo {year} {2017})}\BibitemShut {NoStop}%
	\bibitem [{\citenamefont {Kainaris}\ and\ \citenamefont {Carr}(2015)}]{Carr}%
	\BibitemOpen
	\bibfield  {author} {\bibinfo {author} {\bibfnamefont {N.}~\bibnamefont
			{Kainaris}}\ and\ \bibinfo {author} {\bibfnamefont {S.~T.}\ \bibnamefont
			{Carr}},\ }\href {\doibase 10.1103/PhysRevB.92.035139} {\bibfield  {journal}
		{\bibinfo  {journal} {Phys. Rev. B}\ }\textbf {\bibinfo {volume} {92}},\
		\bibinfo {pages} {035139} (\bibinfo {year} {2015})}\BibitemShut {NoStop}%
	\bibitem [{\citenamefont {Scheller}\ \emph {et~al.}(2014)\citenamefont
		{Scheller}, \citenamefont {Liu}, \citenamefont {Barak}, \citenamefont
		{Yacoby}, \citenamefont {Pfeiffer}, \citenamefont {West},\ and\ \citenamefont
		{Zumb{\"u}hl}}]{zum}%
	\BibitemOpen
	\bibfield  {author} {\bibinfo {author} {\bibfnamefont {C.~P.}\ \bibnamefont
			{Scheller}}, \bibinfo {author} {\bibfnamefont {T.-M.}\ \bibnamefont {Liu}},
		\bibinfo {author} {\bibfnamefont {G.}~\bibnamefont {Barak}}, \bibinfo
		{author} {\bibfnamefont {A.}~\bibnamefont {Yacoby}}, \bibinfo {author}
		{\bibfnamefont {L.~N.}\ \bibnamefont {Pfeiffer}}, \bibinfo {author}
		{\bibfnamefont {K.~W.}\ \bibnamefont {West}}, \ and\ \bibinfo {author}
		{\bibfnamefont {D.~M.}\ \bibnamefont {Zumb{\"u}hl}},\ }\href {\doibase
		10.1103/PhysRevLett.112.066801} {\bibfield  {journal} {\bibinfo  {journal}
			{Phys. Rev. Lett.}\ }\textbf {\bibinfo {volume} {112}},\ \bibinfo {pages}
		{066801} (\bibinfo {year} {2014})}\BibitemShut {NoStop}%
	\bibitem [{\citenamefont {Kammhuber}\ \emph {et~al.}(2017)\citenamefont
		{Kammhuber}, \citenamefont {Cassidy}, \citenamefont {Pei}, \citenamefont
		{Nowak}, \citenamefont {Vuik}, \citenamefont {G{\"u}l}, \citenamefont {Car},
		\citenamefont {Plissard}, \citenamefont {Bakkers}, \citenamefont {Wimmer},\
		and\ \citenamefont {Kouwenhoven}}]{Kammhuber}%
	\BibitemOpen
	\bibfield  {author} {\bibinfo {author} {\bibfnamefont {J.}~\bibnamefont
			{Kammhuber}}, \bibinfo {author} {\bibfnamefont {M.~C.}\ \bibnamefont
			{Cassidy}}, \bibinfo {author} {\bibfnamefont {F.}~\bibnamefont {Pei}},
		\bibinfo {author} {\bibfnamefont {M.~P.}\ \bibnamefont {Nowak}}, \bibinfo
		{author} {\bibfnamefont {A.}~\bibnamefont {Vuik}}, \bibinfo {author}
		{\bibfnamefont {{\"O}.}~\bibnamefont {G{\"u}l}}, \bibinfo {author}
		{\bibfnamefont {D.}~\bibnamefont {Car}}, \bibinfo {author} {\bibfnamefont
			{S.~R.}\ \bibnamefont {Plissard}}, \bibinfo {author} {\bibfnamefont {E.~P.
				A.~M.}\ \bibnamefont {Bakkers}}, \bibinfo {author} {\bibfnamefont
			{M.}~\bibnamefont {Wimmer}}, \ and\ \bibinfo {author} {\bibfnamefont {L.~P.}\
			\bibnamefont {Kouwenhoven}},\ }\href {\doibase 10.1038/s41467-017-00315-y}
	{\bibfield  {journal} {\bibinfo  {journal} {Nature Communications}\ }\textbf
		{\bibinfo {volume} {8}},\ \bibinfo {pages} {478} (\bibinfo {year}
		{2017})}\BibitemShut {NoStop}%
	\bibitem [{\citenamefont {Tsvelik}\ and\ \citenamefont
		{Yevtushenko}(2015)}]{oleg1}%
	\BibitemOpen
	\bibfield  {author} {\bibinfo {author} {\bibfnamefont {A.~M.}\ \bibnamefont
			{Tsvelik}}\ and\ \bibinfo {author} {\bibfnamefont {O.~M.}\ \bibnamefont
			{Yevtushenko}},\ }\href {\doibase 10.1103/PhysRevLett.115.216402} {\bibfield
		{journal} {\bibinfo  {journal} {Phys. Rev. Lett.}\ }\textbf {\bibinfo
			{volume} {115}},\ \bibinfo {pages} {216402} (\bibinfo {year}
		{2015})}\BibitemShut {NoStop}%
	\bibitem [{\citenamefont {Schimmel}\ \emph {et~al.}(2016)\citenamefont
		{Schimmel}, \citenamefont {Tsvelik},\ and\ \citenamefont
		{Yevtushenko}}]{schimmel1}%
	\BibitemOpen
	\bibfield  {author} {\bibinfo {author} {\bibfnamefont {D.~H.}\ \bibnamefont
			{Schimmel}}, \bibinfo {author} {\bibfnamefont {A.~M.}\ \bibnamefont
			{Tsvelik}}, \ and\ \bibinfo {author} {\bibfnamefont {O.~M.}\ \bibnamefont
			{Yevtushenko}},\ }\href {\doibase 10.1088/1367-2630/18/5/053004} {\bibfield
		{journal} {\bibinfo  {journal} {New J. Phys.}\ }\textbf {\bibinfo {volume}
			{18}},\ \bibinfo {pages} {053004} (\bibinfo {year} {2016})}\BibitemShut
	{NoStop}%
	\bibitem [{\citenamefont {Tsvelik}\ and\ \citenamefont
		{Yevtushenko}(2019)}]{oleg2}%
	\BibitemOpen
	\bibfield  {author} {\bibinfo {author} {\bibfnamefont {A.~M.}\ \bibnamefont
			{Tsvelik}}\ and\ \bibinfo {author} {\bibfnamefont {O.~M.}\ \bibnamefont
			{Yevtushenko}},\ }\href {\doibase 10.1103/PhysRevB.100.165110} {\bibfield
		{journal} {\bibinfo  {journal} {Phys. Rev. B}\ }\textbf {\bibinfo {volume}
			{100}},\ \bibinfo {pages} {165110} (\bibinfo {year} {2019})}\BibitemShut
	{NoStop}%
	\bibitem [{\citenamefont {Tsvelik}\ and\ \citenamefont
		{Yevtushenko}(2020)}]{oleg3}%
	\BibitemOpen
	\bibfield  {author} {\bibinfo {author} {\bibfnamefont {A.~M.}\ \bibnamefont
			{Tsvelik}}\ and\ \bibinfo {author} {\bibfnamefont {O.~M.}\ \bibnamefont
			{Yevtushenko}},\ }\href {\doibase 10.1088/1367-2630/ab82bb} {\bibfield
		{journal} {\bibinfo  {journal} {New Journal of Physics}\ }\textbf {\bibinfo
			{volume} {22}},\ \bibinfo {pages} {053013} (\bibinfo {year}
		{2020})}\BibitemShut {NoStop}%
	\bibitem [{\citenamefont {Feldman}\ \emph {et~al.}(2017)\citenamefont
		{Feldman}, \citenamefont {Randeria}, \citenamefont {Li}, \citenamefont
		{Jeon}, \citenamefont {Xie}, \citenamefont {Wang}, \citenamefont {Drozdov},
		\citenamefont {Andrei~Bernevig},\ and\ \citenamefont
		{Yazdani}}]{Feldman2017}%
	\BibitemOpen
	\bibfield  {author} {\bibinfo {author} {\bibfnamefont {B.~E.}\ \bibnamefont
			{Feldman}}, \bibinfo {author} {\bibfnamefont {M.~T.}\ \bibnamefont
			{Randeria}}, \bibinfo {author} {\bibfnamefont {J.}~\bibnamefont {Li}},
		\bibinfo {author} {\bibfnamefont {S.}~\bibnamefont {Jeon}}, \bibinfo {author}
		{\bibfnamefont {Y.}~\bibnamefont {Xie}}, \bibinfo {author} {\bibfnamefont
			{Z.}~\bibnamefont {Wang}}, \bibinfo {author} {\bibfnamefont {I.~K.}\
			\bibnamefont {Drozdov}}, \bibinfo {author} {\bibfnamefont {B.}~\bibnamefont
			{Andrei~Bernevig}}, \ and\ \bibinfo {author} {\bibfnamefont {A.}~\bibnamefont
			{Yazdani}},\ }\href {\doibase 10.1038/nphys3947} {\bibfield  {journal}
		{\bibinfo  {journal} {Nature Physics}\ }\textbf {\bibinfo {volume} {13}},\
		\bibinfo {pages} {286} (\bibinfo {year} {2017})}\BibitemShut {NoStop}%
	\bibitem [{\citenamefont {Desjardins}\ \emph {et~al.}(2019)\citenamefont
		{Desjardins}, \citenamefont {Contamin}, \citenamefont {Delbecq},
		\citenamefont {Dartiailh}, \citenamefont {Bruhat}, \citenamefont {Cubaynes},
		\citenamefont {Viennot}, \citenamefont {Mallet}, \citenamefont {Rohart},
		\citenamefont {Thiaville}, \citenamefont {Cottet},\ and\ \citenamefont
		{Kontos}}]{Desjardins2019}%
	\BibitemOpen
	\bibfield  {author} {\bibinfo {author} {\bibfnamefont {M.~M.}\ \bibnamefont
			{Desjardins}}, \bibinfo {author} {\bibfnamefont {L.~C.}\ \bibnamefont
			{Contamin}}, \bibinfo {author} {\bibfnamefont {M.~R.}\ \bibnamefont
			{Delbecq}}, \bibinfo {author} {\bibfnamefont {M.~C.}\ \bibnamefont
			{Dartiailh}}, \bibinfo {author} {\bibfnamefont {L.~E.}\ \bibnamefont
			{Bruhat}}, \bibinfo {author} {\bibfnamefont {T.}~\bibnamefont {Cubaynes}},
		\bibinfo {author} {\bibfnamefont {J.~J.}\ \bibnamefont {Viennot}}, \bibinfo
		{author} {\bibfnamefont {F.}~\bibnamefont {Mallet}}, \bibinfo {author}
		{\bibfnamefont {S.}~\bibnamefont {Rohart}}, \bibinfo {author} {\bibfnamefont
			{A.}~\bibnamefont {Thiaville}}, \bibinfo {author} {\bibfnamefont
			{A.}~\bibnamefont {Cottet}}, \ and\ \bibinfo {author} {\bibfnamefont
			{T.}~\bibnamefont {Kontos}},\ }\href {\doibase 10.1038/s41563-019-0457-6}
	{\bibfield  {journal} {\bibinfo  {journal} {Nature Materials}\ }\textbf
		{\bibinfo {volume} {18}},\ \bibinfo {pages} {1060} (\bibinfo {year}
		{2019})}\BibitemShut {NoStop}%
	\bibitem [{\citenamefont {J{\"a}ck}\ \emph {et~al.}(2019)\citenamefont
		{J{\"a}ck}, \citenamefont {Xie}, \citenamefont {Li}, \citenamefont {Jeon},
		\citenamefont {Bernevig},\ and\ \citenamefont {Yazdani}}]{yaz}%
	\BibitemOpen
	\bibfield  {author} {\bibinfo {author} {\bibfnamefont {B.}~\bibnamefont
			{J{\"a}ck}}, \bibinfo {author} {\bibfnamefont {Y.}~\bibnamefont {Xie}},
		\bibinfo {author} {\bibfnamefont {J.}~\bibnamefont {Li}}, \bibinfo {author}
		{\bibfnamefont {S.}~\bibnamefont {Jeon}}, \bibinfo {author} {\bibfnamefont
			{B.~A.}\ \bibnamefont {Bernevig}}, \ and\ \bibinfo {author} {\bibfnamefont
			{A.}~\bibnamefont {Yazdani}},\ }\href {\doibase 10.1126/science.aax1444}
	{\bibfield  {journal} {\bibinfo  {journal} {Science}\ }\textbf {\bibinfo
			{volume} {364}},\ \bibinfo {pages} {1255} (\bibinfo {year}
		{2019})}\BibitemShut {NoStop}%
	\bibitem [{\citenamefont {Pfeiffer}\ \emph {et~al.}(1993)\citenamefont
		{Pfeiffer}, \citenamefont {Störmer}, \citenamefont {Baldwin}, \citenamefont
		{West}, \citenamefont {Goñi}, \citenamefont {Pinczuk}, \citenamefont
		{Ashoori}, \citenamefont {Dignam},\ and\ \citenamefont {Wegscheider}}]{CEO}%
	\BibitemOpen
	\bibfield  {author} {\bibinfo {author} {\bibfnamefont {L.}~\bibnamefont
			{Pfeiffer}}, \bibinfo {author} {\bibfnamefont {H.~L.}\ \bibnamefont
			{Störmer}}, \bibinfo {author} {\bibfnamefont {K.~W.}\ \bibnamefont
			{Baldwin}}, \bibinfo {author} {\bibfnamefont {K.~W.}\ \bibnamefont {West}},
		\bibinfo {author} {\bibfnamefont {A.~R.}\ \bibnamefont {Goñi}}, \bibinfo
		{author} {\bibfnamefont {A.}~\bibnamefont {Pinczuk}}, \bibinfo {author}
		{\bibfnamefont {R.~C.}\ \bibnamefont {Ashoori}}, \bibinfo {author}
		{\bibfnamefont {M.~M.}\ \bibnamefont {Dignam}}, \ and\ \bibinfo {author}
		{\bibfnamefont {W.}~\bibnamefont {Wegscheider}},\ }\href {\doibase
		10.1016/0022-0248(93)90746-J} {\bibfield  {journal} {\bibinfo  {journal}
			{Journal of Crystal Growth}\ }\textbf {\bibinfo {volume} {127}},\ \bibinfo
		{pages} {849–857} (\bibinfo {year} {1993})}\BibitemShut {NoStop}%
	\bibitem [{\citenamefont {Mizokuchi}\ \emph {et~al.}(2018)\citenamefont
		{Mizokuchi}, \citenamefont {Maurand}, \citenamefont {Vigneau}, \citenamefont
		{Myronov},\ and\ \citenamefont {De~Franceschi}}]{SiGe}%
	\BibitemOpen
	\bibfield  {author} {\bibinfo {author} {\bibfnamefont {R.}~\bibnamefont
			{Mizokuchi}}, \bibinfo {author} {\bibfnamefont {R.}~\bibnamefont {Maurand}},
		\bibinfo {author} {\bibfnamefont {F.}~\bibnamefont {Vigneau}}, \bibinfo
		{author} {\bibfnamefont {M.}~\bibnamefont {Myronov}}, \ and\ \bibinfo
		{author} {\bibfnamefont {S.}~\bibnamefont {De~Franceschi}},\ }\href {\doibase
		10.1021/acs.nanolett.8b01457} {\bibfield  {journal} {\bibinfo  {journal}
			{Nano Letters}\ }\textbf {\bibinfo {volume} {18}},\ \bibinfo {pages} {4861}
		(\bibinfo {year} {2018})}\BibitemShut {NoStop}%
	\bibitem [{\citenamefont {Tsunetsugu}\ \emph {et~al.}(1997)\citenamefont
		{Tsunetsugu}, \citenamefont {Sigrist},\ and\ \citenamefont {Ueda}}]{KC1}%
	\BibitemOpen
	\bibfield  {author} {\bibinfo {author} {\bibfnamefont {H.}~\bibnamefont
			{Tsunetsugu}}, \bibinfo {author} {\bibfnamefont {M.}~\bibnamefont {Sigrist}},
		\ and\ \bibinfo {author} {\bibfnamefont {K.}~\bibnamefont {Ueda}},\ }\href
	{\doibase 10.1103/RevModPhys.69.809} {\bibfield  {journal} {\bibinfo
			{journal} {Rev. Mod. Phys.}\ }\textbf {\bibinfo {volume} {69}},\ \bibinfo
		{pages} {809} (\bibinfo {year} {1997})}\BibitemShut {NoStop}%
	\bibitem [{\citenamefont {Honner}\ and\ \citenamefont
		{Gul{\'a}csi}(1997)}]{KC2}%
	\BibitemOpen
	\bibfield  {author} {\bibinfo {author} {\bibfnamefont {G.}~\bibnamefont
			{Honner}}\ and\ \bibinfo {author} {\bibfnamefont {M.}~\bibnamefont
			{Gul{\'a}csi}},\ }\href {\doibase 10.1103/PhysRevLett.78.2180} {\bibfield
		{journal} {\bibinfo  {journal} {Phys. Rev. Lett.}\ }\textbf {\bibinfo
			{volume} {78}},\ \bibinfo {pages} {2180} (\bibinfo {year}
		{1997})}\BibitemShut {NoStop}%
	\bibitem [{\citenamefont {Novais}\ \emph
		{et~al.}(2002{\natexlab{a}})\citenamefont {Novais}, \citenamefont {Miranda},
		\citenamefont {{Castro Neto}},\ and\ \citenamefont {Cabrera}}]{KC3}%
	\BibitemOpen
	\bibfield  {author} {\bibinfo {author} {\bibfnamefont {E.}~\bibnamefont
			{Novais}}, \bibinfo {author} {\bibfnamefont {E.}~\bibnamefont {Miranda}},
		\bibinfo {author} {\bibfnamefont {A.~H.}\ \bibnamefont {{Castro Neto}}}, \
		and\ \bibinfo {author} {\bibfnamefont {G.~G.}\ \bibnamefont {Cabrera}},\
	}\href {\doibase 10.1103/PhysRevB.66.174409} {\bibfield  {journal} {\bibinfo
			{journal} {Phys. Rev. B}\ }\textbf {\bibinfo {volume} {66}},\ \bibinfo
		{pages} {174409} (\bibinfo {year} {2002}{\natexlab{a}})}\BibitemShut
	{NoStop}%
	\bibitem [{\citenamefont {Shibata}\ and\ \citenamefont {Ueda}(1999)}]{KC4}%
	\BibitemOpen
	\bibfield  {author} {\bibinfo {author} {\bibfnamefont {N.}~\bibnamefont
			{Shibata}}\ and\ \bibinfo {author} {\bibfnamefont {K.}~\bibnamefont {Ueda}},\
	}\href {\doibase 10.1088/0953-8984/11/2/002} {\bibfield  {journal} {\bibinfo
			{journal} {J. Phys.: Condens. Matter}\ }\textbf {\bibinfo {volume} {11}},\
		\bibinfo {pages} {R1–R30} (\bibinfo {year} {1999})}\BibitemShut {NoStop}%
	\bibitem [{\citenamefont {Xavier}\ and\ \citenamefont {Miranda}(2004)}]{KC6}%
	\BibitemOpen
	\bibfield  {author} {\bibinfo {author} {\bibfnamefont {J.~C.}\ \bibnamefont
			{Xavier}}\ and\ \bibinfo {author} {\bibfnamefont {E.}~\bibnamefont
			{Miranda}},\ }\href {\doibase 10.1103/PhysRevB.70.075110} {\bibfield
		{journal} {\bibinfo  {journal} {Phys. Rev. B}\ }\textbf {\bibinfo {volume}
			{70}},\ \bibinfo {pages} {075110} (\bibinfo {year} {2004})}\BibitemShut
	{NoStop}%
	\bibitem [{\citenamefont {Maciejko}(2012)}]{KC5}%
	\BibitemOpen
	\bibfield  {author} {\bibinfo {author} {\bibfnamefont {J.}~\bibnamefont
			{Maciejko}},\ }\href {\doibase 10.1103/PhysRevB.85.245108} {\bibfield
		{journal} {\bibinfo  {journal} {Phys. Rev. B}\ }\textbf {\bibinfo {volume}
			{85}},\ \bibinfo {pages} {245108} (\bibinfo {year} {2012})}\BibitemShut
	{NoStop}%
	\bibitem [{\citenamefont {Novais}\ \emph
		{et~al.}(2002{\natexlab{b}})\citenamefont {Novais}, \citenamefont {Miranda},
		\citenamefont {{Castro Neto}},\ and\ \citenamefont {Cabrera}}]{KC7}%
	\BibitemOpen
	\bibfield  {author} {\bibinfo {author} {\bibfnamefont {E.}~\bibnamefont
			{Novais}}, \bibinfo {author} {\bibfnamefont {E.}~\bibnamefont {Miranda}},
		\bibinfo {author} {\bibfnamefont {A.~H.}\ \bibnamefont {{Castro Neto}}}, \
		and\ \bibinfo {author} {\bibfnamefont {G.~G.}\ \bibnamefont {Cabrera}},\
	}\href {\doibase 10.1103/PhysRevLett.88.217201} {\bibfield  {journal}
		{\bibinfo  {journal} {Phys. Rev. Lett.}\ }\textbf {\bibinfo {volume} {88}},\
		\bibinfo {pages} {217201} (\bibinfo {year} {2002}{\natexlab{b}})}\BibitemShut
	{NoStop}%
	\bibitem [{\citenamefont {Troyer}\ and\ \citenamefont {W{\"u}rtz}(1993)}]{MC}%
	\BibitemOpen
	\bibfield  {author} {\bibinfo {author} {\bibfnamefont {M.}~\bibnamefont
			{Troyer}}\ and\ \bibinfo {author} {\bibfnamefont {D.}~\bibnamefont
			{W{\"u}rtz}},\ }\href {\doibase 10.1103/PhysRevB.47.2886} {\bibfield
		{journal} {\bibinfo  {journal} {Phys. Rev. B}\ }\textbf {\bibinfo {volume}
			{47}},\ \bibinfo {pages} {2886} (\bibinfo {year} {1993})}\BibitemShut
	{NoStop}%
	\bibitem [{\citenamefont {Zachar}\ \emph {et~al.}(1996)\citenamefont {Zachar},
		\citenamefont {Kivelson},\ and\ \citenamefont {Emery}}]{zach}%
	\BibitemOpen
	\bibfield  {author} {\bibinfo {author} {\bibfnamefont {O.}~\bibnamefont
			{Zachar}}, \bibinfo {author} {\bibfnamefont {S.~A.}\ \bibnamefont
			{Kivelson}}, \ and\ \bibinfo {author} {\bibfnamefont {V.~J.}\ \bibnamefont
			{Emery}},\ }\href {\doibase 10.1103/PhysRevLett.77.1342} {\bibfield
		{journal} {\bibinfo  {journal} {Phys. Rev. Lett.}\ }\textbf {\bibinfo
			{volume} {77}},\ \bibinfo {pages} {1342–1345} (\bibinfo {year}
		{1996})}\BibitemShut {NoStop}%
	\bibitem [{\citenamefont {{Doniach S.}}(1977)}]{doni}%
	\BibitemOpen
	\bibfield  {author} {\bibinfo {author} {\bibnamefont {{Doniach S.}}},\ }\href
	{http://www.sciencedirect.com/science/article/pii/0378436377901905}
	{\bibfield  {journal} {\bibinfo  {journal} {Physica B+C}\ }\textbf {\bibinfo
			{volume} {91}},\ \bibinfo {pages} {231} (\bibinfo {year} {1977})}\BibitemShut
	{NoStop}%
	\bibitem [{\citenamefont {Coleman}(2007)}]{coleman}%
	\BibitemOpen
	\bibfield  {author} {\bibinfo {author} {\bibfnamefont {P.}~\bibnamefont
			{Coleman}},\ }\enquote {\bibinfo {title} {Heavy fermions: Electrons at the
			edge of magnetism},}\ in\ \href {\doibase 10.1002/9780470022184.hmm105}
	{\emph {\bibinfo {booktitle} {Handbook of Magnetism and Advanced Magnetic
				Materials}}}\ (\bibinfo  {publisher} {American Cancer Society},\ \bibinfo
	{year} {2007})\BibitemShut {NoStop}%
	\bibitem [{\citenamefont {Yevtushenko}\ and\ \citenamefont
		{Yudson}(2018)}]{oleg5}%
	\BibitemOpen
	\bibfield  {author} {\bibinfo {author} {\bibfnamefont {O.~M.}\ \bibnamefont
			{Yevtushenko}}\ and\ \bibinfo {author} {\bibfnamefont {V.~I.}\ \bibnamefont
			{Yudson}},\ }\href {\doibase 10.1103/PhysRevLett.120.147201} {\bibfield
		{journal} {\bibinfo  {journal} {Phys. Rev. Lett.}\ }\textbf {\bibinfo
			{volume} {120}},\ \bibinfo {pages} {147201} (\bibinfo {year}
		{2018})}\BibitemShut {NoStop}%
	\bibitem [{Note1()}]{Note1}%
	\BibitemOpen
	\bibinfo {note} {We remind readers that, in the 1D case, spins form a helix
		in a transverse plane. This is reflected by the spin susceptibility $ \langle
		S^+_l S^-_m \rangle \sim e^{\pm 2 i k_F \xi _s |l - m|}, S^\pm = S^x \pm i
		S^y $, which has either \(+2k_F\) or \(-2k_F\) component depending on the
		helix handedness \cite {schimmel1}}\BibitemShut {NoStop}%
	\bibitem [{Note2()}]{Note2}%
	\BibitemOpen
	\bibinfo {note} {We implicitly assume summation over all repeated
		indices.}\BibitemShut {Stop}%
	\bibitem [{Note3()}]{Note3}%
	\BibitemOpen
	\bibinfo {note} {The new band operators \(\protect \operatorname {c}_\pm
		=\protect \frac {1}{\protect \sqrt {2}}\left (\protect \operatorname {c}_1
		\pm \protect \operatorname {c}_2\right )\) are the (anti-) symmetric linear
		combinations of the old operators. The lower + band is thus accompanied by a
		downward shift in energy \(\varepsilon _+=\varepsilon _0-t_\perp \) and vice
		versa.}\BibitemShut {Stop}%
	\bibitem [{\citenamefont {Tsvelik}(2003)}]{tsvelikbook1}%
	\BibitemOpen
	\bibfield  {author} {\bibinfo {author} {\bibfnamefont {A.}~\bibnamefont
			{Tsvelik}},\ }\href
	{https://www.cambridge.org/core/books/quantum-field-theory-in-condensed-matter-physics/9BA1A345E7A6D484FD519005488C44B4}
	{\emph {\bibinfo {title} {{Quantum} {Field} {Theory} in {Condensed} {Matter}
				{Physics}}}}\ (\bibinfo  {publisher} {Cambridge University Press},\ \bibinfo
	{year} {2003})\BibitemShut {NoStop}%
	\bibitem [{Note4()}]{Note4}%
	\BibitemOpen
	\bibinfo {note} {The case where the chemical potential belongs to the upper
		band can be treated analogously.}\BibitemShut {Stop}%
	\bibitem [{SM()}]{SM}%
	\BibitemOpen
	\href@noop {} {}\bibinfo {note} {See the supplemental material for a
		derivation of the effective Lagrangian in the strong tunneling limit; the
		separation of fast and slow spin variables; a derivation of the ground-state
		energy equation in the intermediate tunneling regime; the computation of the
		ground-state energy of gapped 1D Dirac fermions; the renormalization of the
		Luttinger parameter \(K_\psi\); spinless disorder; a derivation of the
		ground-state energy equation in the weak tunneling regime; the effect of
		Dresselhaus spin-orbit interaction; and asymmetrically doped Kondo
		chains}\BibitemShut {NoStop}%
	\bibitem [{\citenamefont {Dresselhaus}(1955)}]{dresselhaus}%
	\BibitemOpen
	\bibfield  {author} {\bibinfo {author} {\bibfnamefont {G.}~\bibnamefont
			{Dresselhaus}},\ }\href {\doibase 10.1103/PhysRev.100.580} {\bibfield
		{journal} {\bibinfo  {journal} {Phys. Rev.}\ }\textbf {\bibinfo {volume}
			{100}},\ \bibinfo {pages} {580–586} (\bibinfo {year} {1955})}\BibitemShut
	{NoStop}%
\end{thebibliography}
